\documentclass[11pt,draftcls,onecolumn]{IEEEtran}
\pdfoutput=1
\usepackage{graphicx, amssymb, stfloats, color, microtype, dsfont, algorithm, algpseudocode, pifont, relsize, multicol, multirow, tabularx, adjustbox}
\usepackage [cmex10]{amsmath}
\usepackage[noadjust]{cite}
\usepackage[hidelinks, bookmarksnumbered=true]{hyperref}
\hypersetup{pdfpagemode=UseNone, pdfstartview=FitH}

\ifCLASSOPTIONcompsoc
\usepackage[caption=false,font=normalsize,labelfont=sf,textfont=sf]{subfig}
\else
\usepackage[caption=false,font=footnotesize]{subfig}
\fi

\usepackage{mathtools}
\newlength{\mymathln}

\IEEEaftertitletext{\vspace{-1\baselineskip}}

\cleardoublepage
\makeatletter
\newsavebox\myboxA
\newsavebox\myboxB
\newlength\mylenA
\newcommand*\xoverline[2][0.6]{%
    \sbox{\myboxA}{$\m@th#2$}%
    \setbox\myboxB\null
    \ht\myboxB=\ht\myboxA%
    \dp\myboxB=\dp\myboxA%
    \wd\myboxB=#1\wd\myboxA
    \sbox\myboxB{$\m@th\overline{\copy\myboxB}$}
    \setlength\mylenA{\the\wd\myboxA}
    \addtolength\mylenA{-\the\wd\myboxB}%
    \ifdim\wd\myboxB<\wd\myboxA%
       \rlap{\hskip 0.5\mylenA\usebox\myboxB}{\usebox\myboxA}%
    \else
        \hskip -0.5\mylenA\rlap{\usebox\myboxA}{\hskip 0.5\mylenA\usebox\myboxB}%
    \fi}
\makeatother

\let\originalleft\left
\let\originalright\right
\renewcommand{\left}{\mathopen{}\mathclose\bgroup\originalleft}
\renewcommand{\right}{\aftergroup\egroup\originalright}

\makeatletter
    \newcommand*{\algrule}[1][\algorithmicindent]{\makebox[#1][l]{\hspace*{.5em}\thealgruleextra\vrule height \thealgruleheight depth \thealgruledepth}}%
\newcommand*{\thealgruleextra}{}
\newcommand*{\thealgruleheight}{.75\baselineskip}
\newcommand*{\thealgruledepth}{.25\baselineskip}

\newcount\ALG@printindent@tempcnta
\def\ALG@printindent{%
    \ifnum \theALG@nested>0
        \ifx\ALG@text\ALG@x@notext
        \else
            \unskip
            \addvspace{-1pt}
            \ALG@printindent@tempcnta=1
            \loop
                \algrule[\csname ALG@ind@\the\ALG@printindent@tempcnta\endcsname]%
                \advance \ALG@printindent@tempcnta 1
            \ifnum \ALG@printindent@tempcnta<\numexpr\theALG@nested+1\relax
            \repeat
        \fi
    \fi
    }%
\usepackage{etoolbox}
\patchcmd{\ALG@doentity}{\noindent\hskip\ALG@tlm}{\ALG@printindent}{}{\errmessage{failed to patch}}
\makeatother

\newbox\statebox
\newcommand{\myState}[1]{%
    \setbox\statebox=\vbox{#1}%
    \edef\thealgruleheight{\dimexpr \the\ht\statebox+1pt\relax}%
    \edef\thealgruledepth{\dimexpr \the\dp\statebox+1pt\relax}%
    \ifdim\thealgruleheight<.75\baselineskip
        \def\thealgruleheight{\dimexpr .75\baselineskip+1pt\relax}%
    \fi
    \ifdim\thealgruledepth<.25\baselineskip
        \def\thealgruledepth{\dimexpr .25\baselineskip+1pt\relax}%
    \fi
    \State #1%
    \def\thealgruleheight{\dimexpr .75\baselineskip+1pt\relax}%
    \def\thealgruledepth{\dimexpr .25\baselineskip+1pt\relax}%
}

\newtheorem{theorem}{Theorem}
\newtheorem{lemma}[theorem]{Lemma}

\makeatletter
\long\def\@makecaption#1#2{\ifx\@captype\@IEEEtablestring%
\footnotesize\begin{center}{\normalfont\footnotesize #1}\\
{\normalfont\footnotesize\scshape #2}\end{center}%
\@IEEEtablecaptionsepspace
\else
\@IEEEfigurecaptionsepspace
\setbox\@tempboxa\hbox{\normalfont\footnotesize {#1.}~~ #2}%
\ifdim \wd\@tempboxa >\hsize%
\setbox\@tempboxa\hbox{\normalfont\footnotesize {#1.}~~ }%
\parbox[t]{\hsize}{\normalfont\footnotesize \noindent\unhbox\@tempboxa#2}%
\else
\hbox to\hsize{\normalfont\footnotesize\hfil\box\@tempboxa\hfil}\fi\fi}
\makeatother

\usepackage[para]{threeparttable}
\usepackage{etoolbox}
\AtEndEnvironment{threeparttable}{\vskip10pt}{}{}
\makeatletter
\patchcmd{\TPT@doparanotes}
  {\TPT@hsize}
  {\TPT@hsize\footnotesize}
  {}
  {}
\makeatother

\begin{document}

\bstctlcite{IEEEexample:BSTcontrol}

\title{Fast Algorithms for Designing Multiple Unimodular Waveforms With Good Correlation Properties}

\author{Yongzhe Li
	\ \ and \ Sergiy A. Vorobyov

\vspace{-.3cm}
\thanks{Y. Li and S. A. Vorobyov are with the Department of Signal Processing and Acoustics, Aalto University, P.O. Box 13000, FI-00076 Aalto, Finland (e-mail: \href{mailto:lyzlyz888@gmail.com}{lyzlyz888@gmail.com}; \href{mailto:yongzhe.li@aalto.fi} {yongzhe.li@aalto.fi}; \href{mailto:svor@ieee.org}{svor@ieee.org}).
Preliminary results for parts of this work have been presented at EUSIPCO'16 and ICASSP'17. 
}
}
\maketitle

\begin{abstract}
In this paper, we develop new fast and efficient algorithms for designing single/multiple unimodular waveforms/codes with good auto- and cross-correlation or weighted correlation properties, which are highly desired in radar and communication systems. The waveform design is based on the minimization of the integrated sidelobe level (ISL) and weighted ISL (WISL) of waveforms. As the corresponding optimization problems can quickly grow to large scale with increasing the code length and number of waveforms, the main issue turns to be the development of fast large-scale optimization techniques. The difficulty is also that the corresponding optimization problems are non-convex, but the required accuracy is high. Therefore, we formulate the ISL and WISL minimization problems as non-convex quartic optimization problems in frequency domain, and then simplify them into quadratic problems by utilizing the majorization-minimization technique, which is one of the basic techniques for addressing large-scale and/or non-convex optimization problems. While designing our fast algorithms, we find out and use inherent algebraic structures in the objective functions to rewrite them into quartic forms, and in the case of WISL minimization, to derive additionally an alternative quartic form which allows to apply the quartic-quadratic transformation. Our algorithms are applicable to large-scale unimodular waveform design problems as they are proved to have lower or comparable computational burden (analyzed theoretically) and faster convergence speed (confirmed by comprehensive simulations) than the state-of-the-art algorithms. In addition, the waveforms designed by our algorithms demonstrate better correlation properties compared to their counterparts.
\end{abstract}


\section{Introduction}
\IEEEpubidadjcol
Waveform/code design, as one of the major problems in radar signal processing \cite{BekkermanTargetDetec06, StoicaProbingDesing, BlumWaveform07, ChenAFWaveform08,  LiSignalSyn, DeMaioPCode09, StoicaNewAlg09, HeMIMO09, DeMaioCodePAPR11, AubryKASig13, BluntWFOverview16}, active sensing \cite{StoicaRFTRSensing12, KriegerMIMOSAR14, WangSARChirp12}, and wireless communications \cite{TSEBook05}, has attracted significant interests over the past several decades \cite{LevanonRadarSignal, DeLongWFClu67, BellWFInfo93, DengPolyphaseWaveform, WicksWF10}. In radar signal processing and active sensing applications, waveform design plays an essential role because ``excellent'' waveforms can ensure higher localization accuracy \cite{BekkermanTargetDetec06}, enhanced resolution capability \cite{LiSignalSyn}, and improved delay-Doppler ambiguity of the potential target \cite{YongzheTBAFJournal}. Moreover, designing waveforms with robustness or adaptiveness is also required for the scenarios with harsh environments that include heterogeneous clutter and/or active jammers \cite{DeLongWFClu67}. In addition, with the advance of multiple-input multiple-output (MIMO) radar  \cite{HaimovichMIMO, LiMIMORadar, VorobyovPMIMO, BlumNoncoherent10}, the problem of joint multiple waveform design is gaining even more importance and tends to grow to large scale.

In order to obtain waveforms with desired characteristics, existing approaches usually resort to manipulations with correlation properties, such as the auto- and cross-correlations between different time lags of waveforms, which serve as the determinant factors for evaluating the quality of designed waveforms \cite{StoicaNewAlg09, HeMIMO09}. Perfect auto- and cross-correlation properties indicate that the emitted waveforms are mutually uncorrelated to any time-delayed replica of them, meaning that the target located at the range bin of interest can be easily extracted after matched filtering, and the sidelobes from other range bins are unable to attenuate it. For example, in the applications such as the spot and barrage noise jamming suppression \cite{StoicaRFTRSensing12} and synthetic aperture radar imaging \cite{LiSignalSyn, HeMIMO09}, waveforms with deep notches towards the time lags (or equivalently, frequency bands), where the jamming or clutter signals are located, are highly desired. On the other hand, it is preferred from the hardware perspective that the waveforms maintain the constant-modulus property, which can reduce the cost of developing advanced amplifiers.  

There is a number of existing waveform design methods based on consideration of the correlation properties \cite{StoicaNewAlg09, HeMIMO09, StoicaProbingDesing, LiSignalSyn, SongWFMaMi15, SongWFMaMi16, ZhaoWF16}. The integrated sidelobe level (ISL), which serves as an evaluation metric for correlation levels of waveforms, or equivalently, the accumulated sidelobes at all time lags, is typically used. If the receiver is fixed to be the matched filter, the waveform design methods are focused on the waveform quality itself. Corresponding waveform designs use the fact that the matched filter can be implemented in terms of the correlation between the waveform and its delayed replica. For example, the method of \cite{StoicaNewAlg09} has proposed to design unimodular waveform in frequency domain using a cyclic procedure based on iterative calculations. A surrogate objective which is minimized by a cyclic algorithm has been introduced, and the methods associated with the ISL and weighted ISL (WISL) minimization therein have been named as CAN and WeCAN, respectively. These methods have been later extended to multiple waveform design in \cite{HeMIMO09}. 

If the receiver is not fixed and therefore has to be jointly optimized with the transmitted waveforms, the focus typically shifts to the so-called mismatched filter (also called instrumental variable filter \cite{StoicaTrRe08}) design at the receiver. Such designs add flexibility as they enable to consider  constraints which are difficult to address otherwise. The receive filter is generally mismatched because it trades off the signal-to-noise ratio in order to improve the signal-to-interference-plus-noise ratio. The corresponding design techniques are typically based on alternating optimization where minimum variance distortionless response (MVDR) filter design is involved. Given the waveforms, finding the optimal MVDR receive filter is typically a computationally simpler problem than the waveform design itself. Therefore, our focus here is the development of computationally efficient algorithms for addressing the core problem of waveform design when the optimal receive filter is the matched filter.    

The computational complexity of algorithms is of crucial importance for the ISL and WISL minimization-based unimodular waveform design problems. Indeed, the corresponding optimization problems can quickly grow to large scale with increasing the code length and number of waveforms. However, such problems are non-convex, while classical large-scale optimization approaches are developed for convex problems with relatively simple objective functions and constraints \cite{Nesterov}. The ISL and WISL objective functions as well as the unimodular constraint to the desired waveforms are in fact complex to deal with and the required accuracy of waveform design is high. 

The aforementioned CAN and WeCAN \cite{StoicaNewAlg09} use a cyclic procedure based on iterative calculations. Although large code length up to several thousands is allowed by CAN and WeCAN, the cost in terms of time for these algorithms can reach several hours or even days when the code length and  required number of waveforms grow large. This is a significant limitation that restricts the design of waveforms in real time. In large-scale optimization, the targeted computational complexity per iteration of an algorithm is linear in dimension of the problem or at most quadratic \cite{Nesterov}. To reduce the computational complexity to a reasonable one, many relevant works resort to the majorization-minimization (MaMi) technique \cite{SongWFMaMi15, SongWFMaMi16, ZhaoWF16, HunterMM04, Naghsh13, VorobyovAFRelay12, SoltanaFracQuadratic17}, which is the basic technique for addressing large-scale and/or non-convex optimization problems with complex objectives \cite{HunterMM04}. For example, \cite{Naghsh13} have dealt with multistatic radar waveform design, where an information-theoretic criterion has been utilized, while \cite{SongWFMaMi15, ZhaoWF16} have been concerned with single- and \cite{SongWFMaMi16} multiple-waveform designs.

In addition to the computational complexity, another important characteristic of large-scale optimization algorithms is the convergence speed/rate \cite{Nesterov}. Although the analytic bounds on the convergence rate may be hard/impossible to derive even for some existing large-scale convex optimization algorithms, the design of algorithms with provably faster convergence speed to tolerance than that of the other algorithms is possible even for non-convex problems considered here. 

In this paper,\footnote{Preliminary results on the ISL and WISL minimization-based designs have been presented in \cite{LiVorobyovEUROSIP16} and \cite{LiVorobyovICASSP17}, respectively.} we focus on the ISL and WISL minimization-based unimodular waveform designs for the matched filter receiver, aiming at developing fast algorithms that reduce the computational complexity and have faster convergence speed than the existing algorithms. The paper is based on a more detailed study of inherent algebraic structures of the objective functions, and concerning MaMi, also designing better majorization functions. The principal goal is to enable the real time waveform design even when the code length and number of waveforms are large. Although our work also employs the MaMi approach, it differs from the previous works in many ways. Different from \cite{SongWFMaMi16}, we formulate the ISL minimization-based unimodular waveform design problem as a non-convex quartic problem by transforming the objective into frequency domain and rewriting it as a norm-based objective. Moreover, we find out and use inherent algebraic structures in WISL expression that enable us to derive the corresponding quartic form into an alternative quartic form which in turns allows to apply the quartic-quadratic transformation. This equivalent form is based on eigenvalue decomposition, which we prove to be unnecessary to compute in our corresponding algorithm. Then the ISL and WISL minimization problems in the form of non-convex quartic optimization are simplified into quadratic forms. It allows us to utilize the MaMi technique where the majorization functions also differ from those of \cite{SongWFMaMi16} and \cite{ZhaoWF16}. Our algorithms have lower or comparable computational burden, faster convergence speed, and demonstrate better correlation properties than the existing state-of-the-art algorithms.

The paper is organized as follows. In Section~\ref{Sec:SigModel}, the signal model and the ISL and WISL minimization-based unimodular waveform design problems are presented. In Section~\ref{Sec:OPTviaMM}, new algorithms for the ISL and WISL minimization problems are detailed. Simulation results are presented in Section~\ref{Sec:Simulation}, while the paper is concluded in Section~\ref{Sec:Conclu}.

\noindent
\emph{{Notations}}:
We use bold uppercase, bold lowercase, and italic letters to denote matrices, column vectors, and scalars, respectively. Notations $\left\| \cdot \right\|$, $ \left\| \cdot \right\|_{\mathrm{F}} $, and $\left| \cdot \right|$ are used for Euclidean norm of a vector, Frobenius norm of a matrix, and absolute value, respectively. Similarly, $\left( \cdot \right)^\ast$, $\left( \cdot \right)^{ \mathrm T }$, and $\left( \cdot \right)^{ \mathrm H }$ stand for conjugate, transpose, and conjugate transpose operations, respectively, while $\mathrm{ vec } \left( \cdot \right)$, $\lambda_{ \mathrm{ max } } \left( \cdot \right)$, and $\mathrm{ max } \left\{ \cdot \right\}$ respectively denote column-wise vectorization of a matrix, largest eigenvalue of a matrix, and maximization operations. Notations $ \lfloor \cdot \rfloor $ and $ \mathrm{mod} ( \cdot, \cdot ) $ stand respectively for the floor function and modulo operation with the first argument being the dividend, while $ \mathcal{T} \left \{  \cdot  \right \} $ denotes the operation of constructing a Hermitian Toeplitz matrix from a vector that coincides with the first column of a matrix and $ \mathrm{ diag } \left\{ \cdot \right\} $ is the operator that picks up diagonal elements from a matrix and writes them into a vector (for matrix argument) or forms a diagonal matrix with main diagonal entries picked up from a vector (for vector argument). In addition, $ \mathrm{tr} \left \{ \cdot \right \} $ stands for the matrix trace, $\Re  \left\{ \cdot \right\}$ stands for the real part of a complex value, $ [ \cdot ]_{i,j} $ denotes the $ (i,j) $th element of a matrix, $\otimes$ and $\odot$ respectively denote Kronecker and Hadamard product operations, $\mathbf{ I }_{ M } $ is the $ M \times M $ identity matrix, and $\mathbf{1}_{M} $ denotes an $ M \times 1 $ vector with all elements equal to 1.

\section{Signal Model and Problem Formulation} \label{Sec:SigModel}
Consider a radar (or communication) system which emits $M$ unimodular and mutually orthogonal waveforms. Each waveform is of code length $P$. Then the whole waveform matrix $ \mathbf{Y} $ of size $ P \times M $ is defined as $\mathbf{Y} \triangleq  \left[  \mathbf{y}_1, \ldots, \mathbf{y}_M  \right]$. Here the $m$th column $\mathbf{ y }_{m}$ corresponds to the $m$th launched waveform. Let the $p$th element of $\mathbf{ y }_{ m }$ be $ y_{ m } ( p ) = e^{ j \psi_{ m } (p) } $ where $\psi_{m} (p)$ is an arbitrary phase value ranging between $-\pi$ and $\pi$. When the number of waveforms $M$ reduces to one, the waveform matrix $ \mathbf{Y} $ shrinks to a column vector. 

The ISL for the set of waveforms $\left\{ y_m (p) \right\}_{m=1,p=1}^{M,P}$ can be expressed as \cite{StoicaNewAlg09}
\begin{align}
  \zeta = \sum_{m=1}^M \;\;\;  \sum_{ \mathclap{\substack{ p=-P+1\\ p \neq 0 } } }^{P-1}   | r_{mm} (p) |^2
                 + \sum_{m=1}^M	\;  \sum_{ \mathclap{ \substack{ m'=1\\ m' \neq m } } }^M  
                           \;\;\;\;   \sum_{ \mathclap{ \substack{ p=-P+1 } } }^{P-1}   | r_{mm'} (p) |^2
                \label{eq:corrLev}
\end{align}
where 
\begin{align}
	r_{mm'} (p)  &\triangleq
	                        \sum_{ \mathclap{\substack{ k=p+1 } } }^P   y_m (k)  y_{m'}^\ast (k-p)
	                        = \big(r_{m'm} (p)\big)^{\ast}
	                        \nonumber
	                        \\
 & 
\qquad
	                        m, m' \in \{ 1,  \ldots,  M  \}; \,
  p \in \left\{ 1, \ldots, P-1  \right\}
\end{align}
is the cross-correlation between the $m$th and $m'$th waveforms at the $p$th time lag. The first term on the right-hand side of \eqref{eq:corrLev} is associated with the auto-correlations, while the second term represents the cross-correlations of the waveforms.

Likewise, the WISL for the waveforms $\left\{ y_m (p) \right\}_{m=1,p=1}^{M,P}$ can be expressed as \cite{StoicaNewAlg09}
\begin{align}
  \zeta_{\mathrm{w}} = \sum_{m=1}^M \;\;\;  \sum_{ \mathclap{\substack{ p=-P+1\\ p \neq 0 } } }^{P-1}   \gamma_{p}^{2}  | r_{mm} (p) |^2
                 + \!\!
                 \sum_{m=1}^M	\;  \sum_{ \mathclap{ \substack{ m'=1\\ m' \neq m } } }^M  
                           \;\;\;\;   \sum_{ \mathclap{ \substack{ p=-P+1 } } }^{P-1}		\gamma_{p}^{2}   | r_{mm'} (p) |^2
                \label{eq:WISLNew}
\end{align}
where $\{ \gamma_p \}_{ p=-P+1 }^{ P-1 }$ are real-valued symmetric weights, i.e., $ \gamma_{p} = \gamma_{-p}, \forall p$, used for controlling the sidelobe levels corresponding to different time lags. If $\gamma_p$ takes zero value, it means that the sidelobe level associated with the $p$th time lag is not considered. If all the controlling weights $\{ \gamma_p \}_{ p=-P+1 }^{ P-1 }$ take the value $1$, then $\zeta_{\mathrm{w}} $ in \eqref{eq:WISLNew} coincides with $\zeta$ in \eqref{eq:corrLev}.

The basic unimodular waveform design problem is then formulated as the synthesize of unimodular and mutually orthogonal waveforms $\left\{ y_m (p) \right\}_{m=1,p=1}^{M,P}$ which have as good as possible auto- and cross-correlation or weighted correlation properties. Using \eqref{eq:WISLNew}, the WISL minimization-based unimodular waveform design problem can be formally expressed as
\begin{align} \label{eq:WISLInit}
   \underset{ \mathbf{ Y } } {  \mathrm{ min } }
   &
   \quad
   \zeta_{\mathrm{w}}
   \nonumber
   \\
   \mathrm{ s.t. }
   &\quad
   \left| y_{m}   (p)  \right| = 1, \; m = 1, \ldots, M; \; p = 1, \ldots, P
 \end{align}
where the constraints ensure the modularity of waveforms, while the orthogonality between waveforms is guaranteed by the objective. Obviously, if all the controlling weights $\{ \gamma_p \}_{ p=-P+1 }^{ P-1 }$ take the value $1$, the problem \eqref{eq:WISLInit} becomes the ISL minimization-based unimodular waveform design problem.

\section{Fast Waveform Design Algorithms} \label{Sec:OPTviaMM}
In this section, we develop fast algorithms for the ISL and WISL minimization-based unimodular waveform designs. The algorithms make use of the MaMi technique and exploit inherent algebraic structures in the objective function \eqref{eq:corrLev}, which allows to reduce the computational complexity. 

\subsection{Fast ISL Minimization-Based Algorithm}
The ISL $\zeta$ in \eqref{eq:corrLev} can be rewritten in the matrix form as
\begin{align}
  \zeta
  =
  \sum_{p = -P+1}^{P-1}
            \left\|
                    \mathbf{R}_p - P \mathbf{I}_M   \delta_p
            \right\|^2
  \label{eq:ISLCAN}
\end{align}
where $\mathbf{R}_p$ is the following $M \times M $ waveform correlation matrix
\begin{align}
  \mathbf{R}_p
  \triangleq
  &
  \left[\!
          \begin{array}{cccc}
            r_{11} (p)    &   r_{12} (p)  &   \ldots    &   r_{1M} (p)  \\
            r_{21} (p)    &   r_{22} (p)  &   \ldots    &   r_{1M} (p)  \\
            \vdots        &   \vdots       &   \ddots   & \vdots        \\
            r_{M1} (p)  &   \ldots       &   \ldots     &   r_{MM} (p)
          \end{array}
  \!\right]
\label{eq:Rp}
\end{align}
and $ \delta_p$ is the Kronecker delta function.

Transforming \eqref{eq:ISLCAN} into frequency domain and performing some derivations, the ISL $\zeta$ can be expressed as \cite{HeMIMO09}
\begin{align}  \label{eq:ISLCANSubII}
  \zeta
  =
  \frac{1}{2P}
  \sum_{ p=1 }^{ 2P }
            \big\|
                    \mathbf{ \tilde{ \tilde{ y } } } ( \omega_p )     
                    \mathbf{ \tilde{ \tilde{ y } } }^{ \mathrm H} ( \omega_p ) 
                    - P \mathbf{ I }_M
            \big\|^2
\end{align}
where 
\begin{align}
\mathbf{ \tilde{ \tilde{ y } } } ( \omega_p )
\triangleq
\sum_{ n=1 }^P  \mathbf{ \tilde{ y } }_n    e^{ -j \omega_p n }, \quad 
	\omega_p \triangleq  \frac{2\pi}{2P}p 
	\label{eq:omegap}
\end{align}
with $\mathbf{ \tilde{ y } }_n$ being the $n$th row of the waveform matrix $ \mathbf{Y} $, i.e.,  
$\mathbf{ \tilde{y} }_n \triangleq \left[ y_1 (n), \ldots, y_M (n) \right]^{ \mathrm T }$.

Expanding the norm in \eqref{eq:ISLCANSubII}, after some elementary algebraic computations, the ISL $\zeta$ can be rewritten as
\begin{align}
  \zeta
  =
  \frac{1}{2P}
  \sum_{ p=1 }^{ 2P }
      \left(
            \big\| \mathbf{ \tilde{ \tilde{ y } } } ( \omega_p )  \big\|^4
            -2P   \big\| \mathbf{ \tilde{ \tilde{ y } } } ( \omega_p )  \big\|^2
            +P^2 M
      \right) .
  \label{eq:ISLCANSubIII}
\end{align}
Moreover, introducing the $MP \times 1$ vectorized version of the waveform matrix $\mathbf{Y}$ as $\mathbf{y} \triangleq \mathrm{ vec } ( \mathbf{ Y } ) = \left[ \mathbf{ y }_1^{ \mathrm T},   \ldots,   \mathbf{ y }_M^{ \mathrm T} \right]^{ \mathrm T}$ and the $ MP \times M$ matrix $\mathbf{A}_p \triangleq \mathbf{I}_M  \otimes \mathbf{a}_p$ with $ \mathbf{a}_{p} $ defined as $\mathbf{ a }_p \triangleq  \big[ 1, e^{ j \omega_p }, \ldots, e^{ j (P-1) \omega_p }  \big]^{ \mathrm T}$ where $p = 1, \ldots, 2P$, and using the facts that $\mathbf{\tilde{\tilde{y}}} (\omega_p) = \mathbf{A}_p^{ \mathrm{H}} \mathbf{y}$ and $\big\| \mathbf{ \tilde{\tilde{y}}} (\omega_p) \big\|^2 = \mathbf{\tilde{ \tilde{y}}}^{\mathrm{H}} (\omega_p) \mathbf{ \tilde{\tilde{y}}} (\omega_p)$, the ISL expression \eqref{eq:ISLCANSubIII} can be further rewritten as 
\begin{align}
  \zeta & = \frac{1}{2P} \sum_{p=1}^{2P} \left( \big( \mathbf{y}^{\mathrm H} \mathbf{A}_p  \mathbf{A }_p^{\mathrm H} \mathbf{y} \big)^2 - 2P \big( \mathbf{y}^{\mathrm H} \mathbf{A}_p  \mathbf{A}_p^{ \mathrm H} \mathbf{y} \big) + P^2 M \vphantom{\big( \mathbf{y}^{\mathrm H}  \mathbf{A}_p  \mathbf{A }_p^{\mathrm H} \mathbf{y} \big)^2} \right) .
  \label{eq:ISLCANFnl}
\end{align}
 
Noticing that $\sum_{p=1}^{2P} \mathbf{A}_p \mathbf{A}_p^{\mathrm H} = 2P \mathbf{ I }_{MP}$ and using the fact that the desired waveforms are orthogonal and have constant modulus, i.e., $\| \mathbf{y} \|^2 = M P$, we can find that 
\begin{align}
  \sum_{p=1}^{2P} \mathbf{y}^{\mathrm H} \mathbf{A}_p \mathbf{A}_p^{\mathrm H} \mathbf{y} = \mathbf{y }^{\mathrm H} \left( \sum_{p=1}^{2P} \mathbf{A}_p \mathbf{A}_p^{\mathrm H} \right) \mathbf{y}
  = 2MP^2 . \label{fact1}
\end{align}
Using \eqref{fact1} and excluding the immaterial optimization terms from \eqref{eq:ISLCANFnl}, the optimization problem \eqref{eq:WISLInit} can be rewritten as
\begin{align}
  \underset{ \mathbf{ y } } {  \mathrm{ min } }
  &
  \quad
  \sum_{ p=1 }^{ 2P }
  \big( \mathbf{ y }^{ \mathrm H }  \mathbf{ A }_p  \mathbf{ A }_p^{ \mathrm H }  \mathbf{ y } \big)^2
  \nonumber
  \\
  \mathrm{ s.t. }
  &
  \quad
  \left| \mathbf{ y }   (p')  \right| = 1, \quad p' = 1, \ldots, MP
  \label{eq:OptCANI}
\end{align}
where the objective function takes a quartic form with respect to $\mathbf{y}$.

Introducing the $MP \times MP$ and $M^2 P^2 \times M^2 P^2$, respectively, matrices $\mathbf{\tilde{Y}} \triangleq \mathbf{y} \mathbf{y}^{\mathrm H}$ and 
\begin{align} \label{eq:Phi}
\boldsymbol{\Phi} \triangleq \sum_{p=1}^{2P} \mathrm{vec} \big( \mathbf{A}_p \mathbf{A}_p^{\mathrm H} \big) \left( \mathrm{vec} \big( \mathbf{A}_p  \mathbf{A }_p^{\mathrm H} \big) \right)^{\mathrm H} 
\end{align}
and using the property that $\mathbf{y}^{\mathrm H} \mathbf{A}_p \mathbf{A }_p^{\mathrm H} \mathbf{y} = \mathrm{tr} \big\{ \mathbf{\tilde{Y}}^{\mathrm H} \mathbf{A}_p  \mathbf{A }_p^{\mathrm H} \big\} = \big( \mathrm{vec} \big( \mathbf{\tilde{Y}} \big) \big)^{\mathrm H} \mathrm{vec} \big( \mathbf{A}_p \mathbf{A}_p^{\mathrm H} \big)$, which follows from the elementary properties of the trace and vectorization operations, the objective function in \eqref{eq:OptCANI} can be transformed from quartic into quadratic form as follows 
\begin{align}
  & \sum_{p=1}^{2P} \big( \mathbf{y}^{\mathrm H} \mathbf{A}_p \mathbf{A}_p^{\mathrm H} \mathbf{y} \big)^2 = \sum_{p=1}^{2P} \big( \mathrm{tr} \big\{ \mathbf{ \tilde{Y}}^{\mathrm H} \mathbf{A}_p  \mathbf{A}_p^{\mathrm H} \big\} \big)^2 \nonumber  \\
    &  = \sum_{p=1}^{2P} \big( \mathrm{vec} \big( \mathbf{\tilde{Y}} \big) \big)^{\mathrm H} \mathrm{vec} \big( \mathbf{A}_p \mathbf{A}_p^{\mathrm H} \big) \left( \mathrm{vec} \big( \mathbf{A }_p \mathbf{A}_p^{\mathrm H} \big) \right)^{\mathrm H} \big( \mathrm{vec } \big( \mathbf{\tilde{Y } } \big) \big)^{\mathrm H} \nonumber \\
    & = \big( \mathrm{vec} \big( \mathbf{\tilde{Y}} \big) \big)^{\mathrm H} \boldsymbol{\Phi} \, \mathrm{vec} \big( \mathbf{ \tilde{Y}} \big) .  \label{eq:TransI}
\end{align}
Therefore, the problem \eqref{eq:OptCANI} can be further rewritten as
\begin{subequations}
\label{eq:OptCANII}
	\begin{alignat}{3}
		\underset{ \mathbf{ \tilde{ Y } } } {  \mathrm{ min } }
		  &
		  \quad
		  \big( \mathrm{ vec } \big( \mathbf{ \tilde{ Y } } \big) \big)^{\mathrm H}
		    \boldsymbol{\Phi} \,
		    \mathrm{ vec } \big( \mathbf{ \tilde{ Y } } \big)
		    \label{eq:OptCANII-a}
		  \\
		  \mathrm{ s.t. }
		  &
		  \quad
		  \mathbf{ \tilde{ Y } }  =   \mathbf{ y } \mathbf{ y }^{ \mathrm H },
		  \quad
		  \left| \mathbf{ y }   (p')  \right| = 1, \; p' = 1, \ldots, MP.
	\end{alignat}
\end{subequations}
Since the objective function \eqref{eq:OptCANII-a} takes a quadratic form, a proper majorized function can be applied. Before applying the majorant to \eqref{eq:OptCANII-a}, we present the following general result that will be used later.

\begin{lemma} \label{th:th1}
	If a real-valued function $f (\mathbf{x})$ with complex variable $\mathbf{x}$ is second-order differentiable, and there is a matrix $\mathbf{G} \succeq 0$ satisfying the generalized inequality $\nabla^{2} f(\mathbf{x}) \preceq \mathbf{G}$ for all $\mathbf{x}$, then for each point $ \mathbf{x}_0$, the following convex quadratic function
	\begin{align}
		g(\mathbf{x}) &= f \left( \mathbf{x}_{0} \right) + \Re \left\{\nabla^{ \mathrm{ H } } f \left( \mathbf{x}_0 \right) \left( \mathbf{x}  - \mathbf{x}_0 \right) \right\} \nonumber \\
		&
		\qquad\qquad\qquad\qquad\quad
		+ \frac{1}{2} \left( \mathbf{x} - \mathbf{x}_{0} \right)^{ \mathrm{ H } } \mathbf{G}	\left( \mathbf{x} - \mathbf{x}_{0} \right)
		\label{eq:gx}
	\end{align}
	majorizes $f (\mathbf{x})$ at $ \mathbf{x}_0$.\footnote{The one-dimension version of Lemma~\ref{th:th1} appears in \cite{Leeuw03} as Theorem 3.1.}
\end{lemma}
\begin{IEEEproof}
Using Taylor's theorem, the second-order expansion of $f (\mathbf{x})$ at the point $ \mathbf{x}_0$ is given as
\begin{align}
	f(\mathbf{x}) &= f \left( \mathbf{x}_0 \right) + \Re \left\{\nabla^{ \mathrm{ H } } f \left( \mathbf{x}_0 \right) \left( \mathbf{x}  - \mathbf{x}_0 \right) \right\} \nonumber \\
			&
			\qquad\qquad\qquad\quad
			+ \frac{1}{2} \left( \mathbf{x} - \mathbf{x}_0 \right)^{ \mathrm{H}} \nabla^2{f \left( \boldsymbol{\xi} \right)  }	\left( \mathbf{x} - \mathbf{x}_0 \right)
\end{align}
where $\boldsymbol{\xi}$ is a point on the line connecting $\mathbf{x}_0$ and $\mathbf{x}$. Due to the fact that $\nabla^2{f \left( \boldsymbol{\xi} \right)} \preceq \mathbf{G}$, the inequality $f(\mathbf{x}) \leq g(\mathbf{x})$ also holds true, where $g ( \mathbf{x} )$ is given by \eqref{eq:gx}.
\end{IEEEproof}

If $f(\mathbf{x} )$ is a quadratic form, i.e., $f(\mathbf{x} ) = \mathbf{x}^{\mathrm{H}} \mathbf{Q} \mathbf{x}$, as it is the case for the objective function in \eqref{eq:OptCANII}, by substituting $\nabla f (\mathbf{x}_0) = 2 \mathbf{Q} \mathbf{x}_0$ in \eqref{eq:gx}, the majorant can be obtained as
\begin{align}
g \left( \mathbf{x}	\right) 
&
= \frac{1}{2} \mathbf{x}^{\mathrm{H}} \mathbf{G} \mathbf{x} + \mathbf{x}_0^{\mathrm{H}}	\left ( \frac{1}{2} \mathbf{G} - \mathbf{Q} \right)	\mathbf{x}_0 \nonumber \\
&
\qquad\qquad\qquad\qquad + 2 \Re \left\{  \mathbf{x}^{ \mathrm{ H } }  \left ( \mathbf{Q} - \frac{1}{2} \mathbf{G} \right )  \mathbf{x}_0 \right\}.
\label{eq:gxQua}
\end{align}

Let $\mathbf{G}$ be the $M^2 P^2 \times M^2 P^2$ identity matrix magnified by the largest eigenvalue of the matrix $\boldsymbol{\Phi}$, i.e., $\mathbf{G} \triangleq \lambda_{\mathrm{max}} ( \boldsymbol{\Phi} ) \mathbf{I}_{ M^2 P^2 }$. For such selection of $\mathbf{G}$, the generalized inequality $\mathbf{G} \succeq \boldsymbol{\Phi}$ is guaranteed to hold. Then using \eqref{eq:gxQua}, the function \eqref{eq:OptCANII-a} can be majorized by the following function
\begin{align}
  & g_1 \big( \mathbf{\tilde{Y}}, \mathbf{ \tilde{Y}}^{(k)} \big) = \tfrac{\lambda_{ \mathrm{max} }  ( \boldsymbol{\Phi} )}{2} \big( \mathrm{vec} \big( \mathbf{\tilde{Y}} \big) \big)^{\mathrm H} \mathrm{vec} \big( \mathbf{\tilde{Y}} \big) \nonumber \\
  & + \big( \mathrm{ vec } \big( \mathbf{ \tilde{ Y } }^{(k)} \big) \big)^{\mathrm H} \left( \tfrac{\lambda_{\mathrm{max}} ( \boldsymbol{\Phi} )}{2} \mathbf{I}_{M^2 P^2} - \boldsymbol{\Phi} \right) \mathrm{ vec} \big( \mathbf{\tilde{Y}}^{(k)} \big) \nonumber \\
  & + 2 \Re \left\{ \big( \mathrm{vec} \big( \mathbf{\tilde{Y}} \big) \big)^{\mathrm H} \left(  \boldsymbol{\Phi} - \tfrac{\lambda_{\mathrm{max}} ( \boldsymbol{\Phi} )}{2} \mathbf{I}_{M^2 P^2}  \right) \mathrm{ vec } \big( \mathbf{ \tilde{ Y } }^{ (k) } \big) \right\}
  \label{eq:MFCANI} 
\end{align}
where the matrix $\mathbf{\tilde{Y}}^{(k)} \triangleq  \mathbf{y}^{(k)} \big( \mathbf{y}^{(k)} \big)^{ \mathrm H}$ is obtained at the $ k $th iteration with $ \mathbf{y}^{(k)} \triangleq \mathrm{vec} \{ \mathbf{Y}^{(k)} \} $ being the vectorized version of the waveform matrix $ \mathbf{Y}^{(k)}$ at iteration $k$.

Using the elementary properties of the Kronecker product and vectorization operations, we can find that 
\begin{align}
\mathrm{ vec } ( \mathbf{ \tilde{ Y } } ) = \mathrm{ vec } \big( \mathbf{ y }   \mathbf{ y }^{ \mathrm H } \big)  =  ( \mathbf{ y }^{ \mathrm T }  \otimes   \mathbf{ I }_{ MP }  )^{ \mathrm H }  \mathbf{ y }.
\label{eq:PropI}
\end{align}
Furthermore, using \eqref{eq:PropI} and the fact that the desired waveforms are orthogonal and unimodular, we obtain
\begin{align}
	\big( \mathrm{vec}  \big ( \mathbf{ \tilde{Y} } \big) \big)^{\mathrm H} \mathrm{vec} \big( \mathbf{\tilde{Y}} \big) = \left\| \mathbf{y} \right\|^4 = M^2 P^2 . \label{eq:ProVecConI}
\end{align}
Moreover, using the definition \eqref{eq:Phi} of the matrix $\boldsymbol{\Phi}$, the maximum eigenvalue of $\boldsymbol{\Phi}$ can be found as
\begin{align}
\lambda_{\mathrm{max}} ( \boldsymbol{\Phi} ) = 2 M P^2 . \label{eq:ProVecConII}
\end{align}
Returning to \eqref{eq:MFCANI} and using the facts \eqref{eq:PropI}--\eqref{eq:ProVecConII}, we can see that the first two terms on the right hand of \eqref{eq:MFCANI} are constant and therefore immaterial for optimization. Thus, ignoring these two terms, the majorization problem for  \eqref{eq:OptCANII} can be written as
\begin{align}
  \underset{ \mathbf{ \tilde{ Y } } } {  \mathrm{ min } }
  & \quad \left( \mathrm{vec} \big( \mathbf{\tilde{Y}} \big) \right)^{\mathrm H} \left( \boldsymbol{\Phi} - M P^2 \mathbf{I}_{M^2 P^2}  \right) \mathrm{vec} \big( \mathbf{ \tilde{Y} }^{ (k)} \big) \nonumber
  \\
  \mathrm{ s.t. }
  & \quad \mathbf{\tilde{Y}} = \mathbf{y} \mathbf{y}^{\mathrm H}; \quad \left| \mathbf{y} (p') \right| = 1, \quad p' = 1, \ldots, MP.
  \label{eq:OptCANIII}
\end{align}

Using the definition \eqref{eq:Phi} and the properties \eqref{eq:PropI} and also
\begin{align}
	\mathrm{vec} \big(\mathbf{A}_p \mathbf{A}_p^{\mathrm H} \big) = \big( \mathbf{A}_p^{\mathrm T}  \otimes \mathbf{I}_{MP} \big)^{\mathrm H} \mathrm{vec} \big(\mathbf{A}_p \big)
	\label{eq:ProII}
\end{align}
the objective function in \eqref{eq:OptCANIII}, denoted hereafter as $\mathrm{obj}^{\mathrm{a}}$, can be expanded as
 \begin{align}
  & \mathrm{obj}^{\mathrm{a}}  = \sum_{p=1}^{2P} \left( \mathbf{y}^{\mathrm H} ( \mathbf{y}^{\mathrm T} \otimes \mathbf{I}_{MP } ) \big( \mathbf{A}_p^{\mathrm T} \otimes \mathbf{I}_{MP} \big)^{\mathrm H} \mathrm{vec} \big( \mathbf{A}_p \big) \right. \nonumber \\
   & \qquad\; \times \left. \left( \mathrm{vec} \big( \mathbf{A}_p \big) \right)^{\mathrm H} \big( \mathbf{ A}_p^{\mathrm T} \otimes \mathbf{I}_{MP} \big) \big( ( \mathbf{y}^{(k)} )^{\mathrm T} \otimes \mathbf{I}_{MP} \big)^{\mathrm H} \mathbf{y}^{(k)} \right) \nonumber \\
   & \qquad\; -  M P^2 \mathbf{y}^{\mathrm H} \left( \mathbf{y}^{\mathrm T} \otimes \mathbf{I}_{MP}  \right) \big( ( \mathbf{y}^{(k)} )^{\mathrm T} \otimes \mathbf{I}_{MP} \big)^{\mathrm H} \mathbf{y }^{(k)}. \label{eq:MFCANDerII}
\end{align}

Applying the mixed-product property of the Kronecker product to the right-hand side of the expansion \eqref{eq:MFCANDerII}, the objective in \eqref{eq:OptCANIII} can be further derived as
\begin{align}
  &
  \mathrm{obj}^{\mathrm{a}} =
  \sum_{p=1}^{ 2P }
  \mathbf{y}^{\mathrm H}
  \left( ( \mathbf{y}^{\mathrm T}   \mathbf{A}_p^\ast )  \otimes  \mathbf{ I }_{ MP } \right)
  \mathrm{vec} \big( \mathbf{A}_p  \big)
  \left( \mathrm{vec} \big( \mathbf{A}_p  \big) \right)^{\mathrm H}
  \big( \big( \mathbf{A}_p^{\mathrm T} 
  \nonumber
  \\
  &
  \;\;\; \times
  ( \mathbf{y}^{(k)} )^\ast \big) \otimes \mathbf{I}_{MP} \big) \mathbf{y}^{(k)} \! - \! M P^2 \mathbf{ y}^{\mathrm H} \big( \mathbf{y}^{\mathrm T} ( \mathbf{y}^{(k)} )^\ast \big) \mathbf{y}^{(k)}.
  \label{eq:MFCANDerIII}
\end{align}
It is straightforward to check that the equality $\left( ( \mathbf{y}^{ \mathrm T} \mathbf{A}_p^\ast ) \otimes \mathbf{I}_{MP} \right) \mathrm{vec} \big( \mathbf{A}_p \big) = \mathbf{A}_p \mathbf{A}_p^{\mathrm H} \mathbf{y}$ holds. Applying this equality to \eqref{eq:MFCANDerIII}, the objective in \eqref{eq:OptCANIII} can be rewritten as
\begin{align}
   &
   \mathrm{obj}^{\mathrm{a}} =
     \sum_{ p=1 }^{ 2P }
     \mathbf{ y }^{ \mathrm H }
     \mathbf{ A }_p  
     \big(
     ( \mathbf{ y }^{ (k) } )^{ \mathrm H }
     \mathbf{ A }_p  \mathbf{ A }_p^{ \mathrm H }
     \mathbf{ y }^{ (k) }
     \big)
     \mathbf{ A }_p^{ \mathrm H }
     \mathbf{ y } - M P^2
     \mathbf{ y }^{ \mathrm H } 
     \nonumber
     \\
  &
  \times \big(
  \mathbf{ y }^{ (k) }  ( \mathbf{ y }^{ (k) } )^{ \mathrm H }  
  \big)
  \mathbf{ y } =
  \mathbf{ y }^{ \mathrm H }
  \big( \mathbf{A}  \boldsymbol{ \Lambda }^{ (k) }   \mathbf{ A }^{ \mathrm H }
        - M P^2 \mathbf{y}^{(k)}  ( \mathbf{ y }^{ (k) } )^{ \mathrm H } \big) \mathbf{ y }
  \label{eq:MFCANDerFnl}
\end{align}
where the  $MP \times 2 MP$ matrix $\mathbf{A}$ and the $2MP \times 2MP$ matrix $\boldsymbol{\Lambda }^{(k)} $ are defined as $\mathbf{A} \triangleq \left[ \mathbf{A}_1, \ldots, \mathbf{A}_{2P} \right]$ and $\boldsymbol{\Lambda}^{(k)} \triangleq \mathrm{ diag } \big\{ \boldsymbol{\mu}^{(k)} \otimes \mathbf{1}_{M}	\big \}$, and the $2P \times 1$ vector $\boldsymbol{\mu}^{(k)}$ is defined as\footnote{In \eqref{eq:mu_k}, $\left | \cdot \right |$ is applied to a matrix argument, which means that the magnitude is found for each element of the matrix, that is, the element-wise magnitude.}
\begin{align}
	\boldsymbol{\mu}^{(k)} 
	 \triangleq  
	\big | \mathbf{\tilde{A}}^{ \mathrm{ H } } \mathbf{Y}^{(k)}  \big |^{2}  \mathbf{1}_{M}
	\label{eq:mu_k}
\end{align}
via the $P \times 2P$ matrix $\mathbf{\tilde{A}} \triangleq \left[ \mathbf{a}_{1}, \ldots, \mathbf{a}_{2P} \right]$. 

Using \eqref{eq:MFCANDerFnl}, the problem \eqref{eq:OptCANIII} can be rewritten as
\begin{align}
  \underset{ \mathbf{ y } } {  \mathrm{ min } }
  &
  \quad
  \mathbf{ y }^{ \mathrm H }
  \big(
        \mathbf{ A }  \boldsymbol{ \Lambda }^{ (k) }   \mathbf{ A }^{ \mathrm H }
        -
         M P^2
        \mathbf{ y }^{ (k) }  ( \mathbf{ y }^{ (k) } )^{ \mathrm H }  
  \big)
  \mathbf{ y }
  \nonumber
  \\
  \mathrm{ s.t. }
  &
  \quad
  \left| \mathbf{ y }   (p')  \right| = 1, \; p' = 1, \ldots, MP 
  \label{eq:OptCANIV}
\end{align}
where the objective function takes a quadratic form, to which the majorant \eqref{eq:gxQua} can be applied again. Substituting the $2MP \times 2MP$ matrix $\mathbf{G}$, defined as $\mathbf{G} \triangleq \mu_{\mathrm{max}}^{(k)} \mathbf{A} \mathbf{A}^{\mathrm H}$, into \eqref{eq:gxQua}, we find that \eqref{eq:MFCANDerFnl} can be majorized by the following function 
\begin{align}
&
g_2 \big( \mathbf{ y }, \mathbf{ y }^{ (k) } \big)
=
\tfrac{1}{2}
\mu_{ \mathrm{ max } }^{ (k) }
\mathbf{ y }^{ \mathrm H } \mathbf{ A } \mathbf{ A }^{ \mathrm H } \mathbf{ y }
+
( \mathbf{ y }^{ (k) } )^{ \mathrm H }
\big(
M P^2 \mathbf{ y }^{ (k) }  ( \mathbf{ y }^{ (k) } )^{ \mathrm H }
\nonumber
\\
&
\quad
-
\mathbf{ A }  ( \boldsymbol{\Lambda}^{ (k) }  
- \tfrac{1}{2} \mu_{ \mathrm{ max } }^{ (k) }  \mathbf{ I }_{ 2MP } )  \mathbf{ A }^{ \mathrm H } 
\big)
\mathbf{ y }^{ (k) }
+
2 \Re
\big\{
\mathbf{ y }^{ \mathrm H } 
\big(
\mathbf{ A }  ( \boldsymbol{\Lambda}^{ (k) }  
\nonumber
\\
&
\quad
- 
\tfrac{1}{2} \mu_{ \mathrm{ max } }^{ (k) }  
\mathbf{ I }_{ 2MP } ) 
\mathbf{ A }^{ \mathrm H } 
- M P^2 \mathbf{ y }^{ (k) }  ( \mathbf{ y }^{ (k) } )^{ \mathrm H }
\big)
\mathbf{ y }^{ (k) }
\big\}
\label{eq:MFCANII}
\end{align}
where $\mu_{\mathrm{max}}^{(k)}$ is the largest element of $\boldsymbol{ \Lambda }^{ (k) }$, equivalently, $\mu_{\mathrm{max}}^{(k)} \triangleq \mathrm{max} \left\{ \boldsymbol{\mu}^{(k)} \right \}$. This scaling factor guaranties that the generalized inequality $\mathbf{G} \succeq \mathbf{A} \boldsymbol{\Lambda }^{(k)} \mathbf{A}^{\mathrm H}$ holds.

Noticing that $\mathbf{A} \mathbf{A}^{\mathrm H} = \sum_{p=1}^{2P} \mathbf{A}_p \mathbf{A}_p^{\mathrm H} = 2P \mathbf{I}_{MP}$ and using the fact that the desired waveforms are orthogonal and have constant modulus, i.e., $\mathbf{y}^{\mathrm H} \mathbf{y} = ( \mathbf{y}^{(k)} )^{\mathrm H} \mathbf{y }^{(k)} = \left\| \mathbf{y} \right\|^2 = MP$, we can see that the first two terms in \eqref{eq:MFCANII} are constant, and hence, immaterial for optimization. Ignoring these terms, the optimization problem \eqref{eq:OptCANIV} can be further majorized by the following problem
\begin{align}
  \underset{ \mathbf{ y } } {  \mathrm{ min } }
  &
  \quad
  \mathbf{ y }^{ \mathrm H }	
  \left(
  \mathbf{ A }  ( \boldsymbol{\Lambda}^{ (k) }  
  - \tfrac{1}{2} \mu_{ \mathrm{ max } }^{ (k) }  \mathbf{ I }_{ 2MP } )  \mathbf{ A }^{ \mathrm H } 
  - M^2 P^3 
  \mathbf{I}_{MP}
  \right)
  \mathbf{y}^{(k)}
  \nonumber
  \\
  \mathrm{ s.t. }
  &
  \quad
  \left| \mathbf{ y }   (p')  \right| = 1, \; p' = 1, \ldots, MP.
  \label{eq:OptCANV}
\end{align}

Using again the fact that the desired waveforms have constant modulus, the problem \eqref{eq:OptCANV} can be equivalently rewritten as
\begin{align}
\underset{ \mathbf{ Y } } {  \mathrm{ min } }
&
\quad
\big\|
\mathbf{ Y } -   \mathbf{ T }^{ (k) }   \mathbf{Y}^{(k)}
\big\|
\nonumber
\\
\mathrm{ s.t. }
&
\quad
\left| [ \mathbf{ Y } ]_{m,p}  \right| = 1, \; m = 1, \ldots, M; \, p = 1, \ldots, P
\label{eq:OptCANEqv}
\end{align}
where the $P \times P$ matrix $\mathbf{T}^{(k)} \triangleq \mathcal{T} \left\{ \mathbf{v}^{(k)} \right \} $ is a Hermitian Toeplitz matrix constructed from the $P \times 1$ vector $\mathbf{v}^{(k)}
\triangleq - \mathbf{\tilde{A}}	\left( \boldsymbol{\mu}^{(k)} - \tfrac{1}{2} \big( \mu_{\mathrm{max} }^{(k)} + M^2 P^2 \big) \mathbf{1}_{2P}	\right)$. The problem \eqref{eq:OptCANEqv} has the following closed-form solution
\begin{align}
[ \mathbf{ Y } ]_{m,p} = \exp \left\{ j \cdot \mathrm{ arg } \left( \left[ \mathbf{ T }^{ (k) } \mathbf{Y}^{(k)} \right]_{m,p}  \right)   \right\},
\; \forall m, \, \forall p.
\label{eq:SolCAN}
\end{align}

\begin{figure}[!t]%
  \vspace{-7pt}
  \begin{algorithm}[H]
    \caption{ISL Minimization-Based Algorithm}
    \label{CANMM}
    \begin{algorithmic}[1]
    \myState {$k \leftarrow 0$, $ \mathbf{ Y }\leftarrow$ unimodular sequence matrix with random phases.}
      \Repeat
      \Procedure {ISLMaMi}{}{$\big(\mathbf{ Y }^{ ( k )  } \big)$}
      \myState
        {$\boldsymbol{\mu}^{(k)} 
        	 =  
        	\big | \mathbf{\tilde{A}}^{ \mathrm{ H } } \mathbf{Y}^{(k)}  \big |^{2}  \mathbf{1}_{M}$}
      \myState{$ 	\mathbf{v}^{(k)}
      	=
      	-	\mathbf{\tilde{A}}	\left ( \boldsymbol{\mu}^{(k)} - \tfrac{1}{2}	\big ( \mu_{ \mathrm{ max } }^{ (k) }  + M^2 P^2 \big ) 	\mathbf{1}_{2P}	\right ) $}
      	\myState{$ \mathbf{T}^{(k)} = \mathcal{T} \left \{ \mathbf{v}^{(k)} \right \} $}
                \myState{$
                  \begin{aligned}
                    [ \mathbf{ Y } ]_{m,p} =
                     e^{ j \cdot \mathrm{ arg } \left( \left[ \mathbf{ T }^{ (k) } \mathbf{Y}^{(k)} \right]_{m,p}  \right)   },
                    \quad \forall m, \, \forall p 
                  \end{aligned}$}
        \myState{$ k \leftarrow k+1$}
          \EndProcedure
      \Until convergence
    \end{algorithmic}
  \end{algorithm}
  \vspace{-20pt}
\end{figure}

Finally, according to the MaMi procedure and using the closed-form solution \eqref{eq:SolCAN} to the majorization problem, the ISL minimization-based unimodular waveform design algorithm is summarized in Algorithm~\ref{CANMM}. There exist accelerated schemes for MaMi, such as the squared iterative method (SQUAREM) of \cite{Varadhan08}, which can be straightforwardly applied to speed up Algorithm~\ref{CANMM}. The SQUAREM scheme is an extension of the scalar Steffensen type method \cite{Henrici64}, \cite{Steff} to vector fixed-point iteration empowered with the idea of ``squaring'' \cite{Raydan02}. It is an ``off-the-shelf'' acceleration method that requires nothing extra to the parameter updating rules of an original algorithm, except possibly the computationally cheap projection to feasibility set, and it is guaranteed to converge \cite{ZhaoWF16, Varadhan08}. 

Different stopping criteria can be employed in Algorithm~\ref{CANMM}. For example, it can be the absolute ISL difference between the current and previous iterations normalized by the initial ISL, 
or it can be the norm of the difference between the waveform matrices obtained at the current and previous iterations. 

In terms of the per iteration computational complexity of Algorithm~\ref{CANMM}, the straightforward calculation of $\boldsymbol{\mu}^{(k)}$ according to \eqref{eq:mu_k} requires $2MP(P+1)$ operations, the calculation of $ \mathbf{v}^{(k)}$ costs $2P^2$ operations, while the computational burden of the matrix to matrix product $\mathbf{T}^{(k)} \mathbf{Y}^{(k)}$ in \eqref{eq:SolCAN} is $MP^2$ operations. Therefore, the total computational complexity is $(3M+2) P^2 + 2MP$ operations. However, $\boldsymbol{\mu}^{(k)}$ and $\mathbf{v}^{(k)}$ can be computed by means of the fast Fourier transform (FFT) at the order of complexity $\mathcal{O}(MP\log{ P})$ and $\mathcal{O}(P\log{P})$, respectively. Similarly, using the Toeplitz structure of $\mathbf{T}^{(k)}$, the product  $\mathbf{T}^{(k)} \mathbf{Y}^{(k)}$ can also be calculated at a reduced complexity  $\mathcal{O}(M P\log{P})$, which is the highest in Algorithm~\ref{CANMM}. Thus, the order of complexity of Algorithm~\ref{CANMM} is $\mathcal{O}(M P\log{P})$, which is nearly linear in the dimension of the problem,
 as required in large-scale optimization. 

\subsection{Fast WISL Minimization-Based Algorithm}
The WISL in \eqref{eq:WISLNew} can be written in a matrix form as
\begin{align} \label{eq:WISLInitNew}
	\zeta_{\mathrm{w}}
	=
	\gamma_{0}^{2} 	
	\left \|	\mathbf{R}_{0} - P \mathbf{I}_{M}	\right \|^{2}
	+
	\sum_{ {\substack{ p=-P+1\\ p \neq 0 } } }^{P-1}  
	\gamma_{p}^{2}	\left \|	\mathbf{R}_{p}	\right \|
\end{align}
where $ \mathbf{R}_{p}, \, p \in \{ -P+1, \ldots, P-1 \} $ are defined in \eqref{eq:Rp}.

In the frequency domain, \eqref{eq:WISLInitNew} can be expressed as \cite{HeMIMO09}
\begin{align}
  \zeta_{\mathrm{w}} = \frac{1}{2P} \sum_{p=1}^{2P} \big\| \boldsymbol{\Psi} ( \omega_p ) - \gamma_0 P \mathbf{ I }_M \big\|^2 \label{eq:WISLFre}
\end{align}
where $\omega_p$ is defined in \eqref{eq:omegap} and
\begin{align}
  \boldsymbol{\Psi} ( \omega_p ) \triangleq \sum_{p = -P+1}^{P-1} \gamma_{p} \mathbf{R}_p e^{-j \omega_p  n}
  \label{eq:SpecMWeCAN}
\end{align}
is the weighted spectral density matrix. 

Let us also define the $P \times P$ Toeplitz matrix constructed by the weights $\{ \gamma_p \}_{ p=-P+1 }^{ P-1 }$ as follows
\begin{align}
  \boldsymbol{ \Gamma }
  \triangleq
  \left[\!\!\!
  \begin{array}{cccc}
      \gamma_0          &   \gamma_1    &     \ldots              &     \gamma_{ P-1 } \\
      \gamma_{-1}          &   \gamma_0    &     \ddots             &     \vdots                 \\
      \vdots                 &   \ddots           &     \ddots             &      \gamma_1          \\
      \gamma_{ -P+1 }  &   \ldots            &      \gamma_{-1}     &       \gamma_0
  \end{array}
  \!\!\!\right] .
  \label{eq:GaMatrix}
\end{align}
Then the matrix $\boldsymbol{\Psi} ( \omega_p )$ in \eqref{eq:SpecMWeCAN} can be rewritten in the vector-matrix form as
\begin{align}
  \boldsymbol{\Psi} \left( \omega_p \right) & = \mathbf{Y}^{\mathrm H} \left( \mathrm{diag} \left\{ \mathbf{a}_p \right\} \right)^{\mathrm H} \boldsymbol{\Gamma} \mathrm{diag}  \left\{  \mathbf{ a }_p  \right\} \mathbf{Y} \nonumber \\
  & = \mathbf{ Y }^{ \mathrm H }  \left( (  \mathbf{ a }_p   \mathbf{ a }_p^{ \mathrm H }  )  \odot   \boldsymbol{ \Gamma } \right) \mathbf{ Y }.
  \label{eq:SpecMSubWeCAN}
\end{align}
Substituting \eqref{eq:SpecMSubWeCAN} into \eqref{eq:WISLFre}, we arrive to the following WISL expression
\begin{align}
  \zeta_{  \mathrm{ w }  } 
  =
  \frac{1}{2P}
  \sum_{ p=1 }^{ 2P } 
      \big\|
              \mathbf{ Y }^{ \mathrm H }  \left( (  \mathbf{ a }_p   \mathbf{ a }_p^{ \mathrm H }  )  \odot   \boldsymbol{ \Gamma }   \right)    \mathbf{ Y } - \gamma_0 P \mathbf{ I }_M
      \big\|^2.
  \label{eq:ISLWeCANSubII}
\end{align}
Expanding the squared norm in the sum of \eqref{eq:ISLWeCANSubII} yields 
\begin{align}
  \zeta_{ \mathrm{ w } } 
  &
  =
  \frac{1}{2P}
  \sum_{ p=1 }^{ 2P } 
  \left(
          \big\|  \mathbf{ Y }^{ \mathrm H }  \left( (  \mathbf{ a }_p   \mathbf{ a }_p^{ \mathrm H }  )  \odot   \boldsymbol{ \Gamma }   \right)    \mathbf{ Y }   \big\|^2
          +
          \gamma_0^2  M  P^2 
          \right.
  \nonumber
  \\
  &
  \qquad\qquad\qquad\;\,
  \left. \vphantom{ \big\|  \mathbf{ Y }^{ \mathrm H } \big\|^2 }
          -
          2 \gamma_0  P  \mathrm{ tr } \left\{  \mathbf{ Y }^{ \mathrm H }  \left( (  \mathbf{ a }_p   \mathbf{ a }_p^{ \mathrm H }  )  \odot   \boldsymbol{ \Gamma }   \right)    \mathbf{ Y }  \right\}
  \right).
  \label{eq:ISLWeCANSubIII}
\end{align}

Using the facts that the desired waveforms are orthogonal and unimodular, i.e., $\mathrm{tr} \left\{ \mathbf{Y}^{\mathrm{H}} \mathbf{Y} \right\} = \left\| \mathbf{Y} \right\|^2 = MP$, and also that  $\sum_{p=1}^{2P} \mathbf{a}_p \mathbf{a}_p^{ \mathrm H} = 2P \mathbf{I}_{P}$, we find that 
\begin{align}
  &
  \sum_{ p=1 }^{ 2P } 
  \mathrm{ tr } \big\{  \mathbf{ Y }^{ \mathrm H }  \left( (  \mathbf{ a }_p   \mathbf{ a }_p^{ \mathrm H }  ) 
  \!
   \odot
   \!
      \boldsymbol{ \Gamma }   \right)    \mathbf{ Y }  \big\}
  =
  \mathrm{ tr  } \left\{  
  \!
  \mathbf{ Y }^{ \mathrm H }  \Big( \! \Big(  \sum_{ p=1 }^{ 2P }  \mathbf{ a }_p   \mathbf{ a }_p^{ \mathrm H }  \Big) \!  \odot  \!  \boldsymbol{ \Gamma }   \Big)    \mathbf{ Y }  
  \!
  \right\}
  \nonumber
  \\
  &
  =
  2 P
  \mathrm{ tr } \left\{  \mathbf{ Y }^{ \mathrm H }  \left( \mathbf{ I }_P  \odot   \boldsymbol{ \Gamma }   \right)    \mathbf{ Y }  \right\}
  =
  2 \gamma_0	P	\left\| \mathbf{ Y } \right\|^2
  =
  2 \gamma_0 M P^2.
  \label{eq:WISLPro}
\end{align}
Therefore, the second and third terms of \eqref{eq:ISLWeCANSubIII} are constant and immaterial for optimization. With this observation, the WISL minimization problem \eqref{eq:WISLInit} can be rewritten as
\begin{subequations}
	\label{eq:WISLFreOpt}
	\begin{alignat}{2}
		   \underset{ \mathbf{ Y } } {  \mathrm{ min } }
		   &
		   \quad
		   		\sum_{ p=1 }^{ 2P }
		   					\big\|  \mathbf{ Y }^{ \mathrm H }  \left( (  \mathbf{ a }_p   \mathbf{ a }_p^{ \mathrm H }  )  \odot   \boldsymbol{ \Gamma }   \right)    \mathbf{ Y }   \big\|^2
			\label{eq:WISLFreOptObj}
		   \\
		   \mathrm{ s.t. }
		   &\quad
		   \left| y_{m}   (p)  \right| = 1, \; m = 1, \ldots, M; \; p = 1, \ldots, P.
	\end{alignat}
\end{subequations}

The Hadamard product of two matrices appears under the Frobenius norm in \eqref{eq:WISLFreOptObj}, and the resulting matrix there is complex. As a result, we cannot arrive to a proper quartic form with respect to $\mathbf{y}$ by directly expanding the squared norm of \eqref{eq:WISLFreOptObj}. Instead, we need to convert it into a proper one. Towards this end, let us consider the eigenvalue decomposition of  $\boldsymbol{\mathbf{\Gamma}}$, which in general may be indefinite and can be expressed as
\begin{align}
	 \boldsymbol{ \mathbf{ \Gamma }} = \sum_{ k=1 }^{ K }	\lambda_k  \mathbf{ q }_k \mathbf{ q }_k^{ \mathrm{ H } }	
	 =
	 \sum_{ k=1 }^{K} \mathbf{ u }_k \mathbf{ v }_k^{ \mathrm{ H } }
	 \label{eq:EigdecGamma}
\end{align}
where $\lambda_k$ and $ \mathbf{q}_k $ are the $k$th eigenvalue and eigenvector, respectively, $\mathbf{u}_k \triangleq \sqrt{\lambda_k} \mathbf{q}_k $, $ \mathbf{v}_{k} $ equals $-\mathbf{u}_{k}$ when $ \lambda_{k} $ is negative, otherwise it is the same as $ \mathbf{u}_{k} $, and $K$ is the rank of $\boldsymbol{\mathbf{\Gamma}}$. Substituting \eqref{eq:EigdecGamma} into \eqref{eq:WISLFreOptObj} and expanding the Frobenius norm, the objective function \eqref{eq:WISLFreOptObj}, called hereafter as $\mathrm{obj}^{\mathrm{b}}$, can be rewritten as
\begin{align}
  \mathrm{obj}^{\mathrm{b}} = \sum_{p=1}^{2P} \sum_{k=1}^{K}  \sum_{k'=1}^{K} \left| \left( \mathbf{v}_{k'} \odot \mathbf{a}_{p} \right)^{\mathrm {H}} \mathbf{Y}^{\mathrm{T}} \mathbf{Y}^{\ast} \left( \mathbf{u}_{k}	\odot \mathbf{a}_{p}\right) \right |^2.
  \label{eq:WISLOptSubI}
\end{align}
Applying the property $\mathbf{Y}^{\ast} \left( \mathbf{u}_{k} \odot \mathbf{a}_{p} \right) = \mathbf{y}^{\mathrm{H}} \left( \mathbf{I}_{M} \otimes \left( \mathbf{a}_{p} \odot \mathbf{u}_{k} \right)	\right)$ (also holds when $ \mathbf{u}_{k} $ is replaced by $ \mathbf{v}_{k'} $) to \eqref{eq:WISLOptSubI} together with the mixed-product property of the Kronecker product, the objective function \eqref{eq:WISLFreOptObj} can be rewritten as
\begin{align}
	{\rm obj} \nonumber 
		&
		=
		\sum_{p=1}^{2P} \sum_{k=1}^{K}  \sum_{k'=1}^{K}
				\left |
				\mathbf{y}^{ \mathrm{ H } }
				\left(
				\mathbf{I}_{M}
				\otimes
				\left( 
				\left(	\mathbf{a}_{p}	\mathbf{a}_{p}^{ \mathrm{ H } }		\right)
					\odot  \left( \mathbf{u}_{k}	 \mathbf{v}_{k'}^{ \mathrm{ H } }	\right)	
				\right)	
				\right)
				\mathbf{y}
				\right |^2
				\\
				&
				=
				\sum_{p=1}^{2P} \sum_{k=1}^{K}  \sum_{k'=1}^{K}
				\left(
								\mathbf{y}^{ \mathrm{ H } }
								\left(
								\left(	\mathbf{A}_{p}	\mathbf{A}_{p}^{ \mathrm{ H } }		\right)
									\odot  
									\boldsymbol{\Gamma}^{ \mathrm{ real } }_{kk'}
								\right)
								\mathbf{y}
				\right)^2
				\nonumber
				\\
				&
				\qquad\qquad\qquad\quad\;
				+
				\left(
												\mathbf{y}^{ \mathrm{ H } }
												\left(
												\left( 
												\mathbf{A}_{p}	\mathbf{A}_{p}^{ \mathrm{ H } }		\right)
													\odot  
													\boldsymbol{\Gamma}^{ \mathrm{ img } }_{kk'}
												\right)
												\mathbf{y}
								\right)^2 \label{obj3WISL}
\end{align}
where the $MP \times MP$ Hermitian matrices $\boldsymbol{\Gamma}^{\mathrm{real}}_{kk'}$ and $\boldsymbol{\Gamma}^{ \mathrm{img}}_{kk'} $ are defined as $\boldsymbol{ \Gamma }_{kk'}^{\mathrm{real}}
\triangleq \mathbf{I}_{M} \otimes \left( \mathbf{u}_{k} \mathbf{v}_{k'}^{\mathrm{H}} + \mathbf{v }_{k'} \mathbf{u}_{k}^{\mathrm{H}} \right) / 2$ and $\boldsymbol{ \Gamma }_{kk'}^{\mathrm{img}}  \triangleq \mathbf{I}_{M} \otimes  \left( \mathbf{u}_{k} \mathbf{v}_{k'}^{\mathrm{H}} - \mathbf{v}_{k'} \mathbf{u}_{k}^{\mathrm{H}} \right)^{\ast} /2$. Substituting \eqref{obj3WISL} to \eqref{eq:WISLFreOpt}, the WISL minimization problem becomes
\begin{subequations}
	\label{eq:OptWeCANIConstr2nd}
	\begin{alignat}{3}
		\underset{ \mathbf{ y } } {  \mathrm{ min } }
				&
			  \quad
					\sum_{ p=1 }^{ 2P }
						\sum_{ k=1 }^{ K }
						\sum_{ k'=1 }^{ K }
								\left(
												\mathbf{ y }^{ \mathrm{ H } }
												\left(
															\big(
																			\mathbf{ A }_p	\mathbf{ A }_p^{ \mathrm{ H } }
															\big)
															\odot
															\boldsymbol{ \Gamma }_{kk'}^{\mathrm{real}}
												\right)
												\mathbf{ y }
								\right)^2
		\nonumber
		\\
		&
		\qquad\qquad\quad\quad\quad\;
		+
		\left(
														\mathbf{ y }^{ \mathrm{ H } }
														\left(
																	\big(
																					\mathbf{ A }_p	\mathbf{ A }_p^{ \mathrm{ H } }
																	\big)
																	\odot
																	\boldsymbol{ \Gamma }_{kk'}^{\mathrm{img}}
														\right)
														\mathbf{ y }
										\right)^2
		\label{eq:OptWeCANObj}
	  \\
	  \mathrm{ s.t. }
	  &
	  \quad
	  \left| \mathbf{ y } (p')  \right| = 1, \quad p' = 1, \ldots, MP .
	\end{alignat}
	\label{eq:OptWeCANI}
\end{subequations}
The objective function \eqref{eq:OptWeCANObj} takes a proper quartic form with respect to $\mathbf{y}$ that enables us to design an algorithm based on the MaMi approach. 

By means of the trace and vectorization operations for matrices, and similar to the previous subsection, we can transform \eqref{eq:OptWeCANObj}, denoted for brevity as $\mathrm{obj}^{\mathrm{c}}$, into the following form
\begin{align}
	\mathrm{obj}^{\mathrm{c}} & = \sum_{p=1}^{2P} \left( \mathrm{tr} \left\{ \mathbf{\tilde{Y}}^{\mathrm{H}} \left( \big( \mathbf{A}_p \mathbf{A}_p^{\mathrm{H}} \big) \odot \boldsymbol{\Gamma}_{kk'}^{\mathrm{real}} \right) \right\} \right)^2 \nonumber \\
		& \qquad\qquad + \left( \mathrm{tr} \left\{ \mathbf{\tilde{Y}}^{\mathrm{H}} \left( \big( \mathbf{A}_p	\mathbf{A}_p^{ \mathrm{H}} \big) \odot \boldsymbol{\Gamma}_{kk'}^{\mathrm{img}} \right) \right\} \right)^2 \nonumber \\
		& = \big( \mathrm{vec} \big( \mathbf{\tilde{Y}} \big) \big)^{\mathrm{H}} \boldsymbol{\tilde{ \Phi}} \, \mathrm{vec} \big( \mathbf{\tilde{Y}} \big) \label{eq:WISLObjSubI}
\end{align}
where $\mathbf{\tilde{Y}} \triangleq \mathbf{y} \mathbf{y}^{\mathrm{H}}$ has been defined before, $\boldsymbol{ \tilde{\Phi}}$ is the $ M^2 P^2 \times M^2 P^2 $ matrix defined as
\begin{align}
	\boldsymbol{ \tilde{ \Phi } } 
	\triangleq
		\boldsymbol{ \bar{ \Phi }}	\odot	\boldsymbol{ \bar{ \Gamma } }
		\label{eq:PhiTilDef}
\end{align}
with
\begin{align}
 \boldsymbol{ \bar{ \Phi }}	
 &
 \triangleq	
 \sum_{p=1}^{2P}	
	\mathrm{vec} \big( \mathbf{A}_p	\mathbf{A}_p^{ \mathrm{ H } } \big)
	\left( \mathrm{vec} \big( \mathbf{A}_p\mathbf{A}_p^{\mathrm{H}} \big) \right)^{\mathrm{H}} \\
	\boldsymbol{ \bar{ \Gamma } }
	&
	\triangleq
		\sum_{k=1}^{K}
			\sum_{k'=1}^{K}
		\mathrm{ vec } \big( \boldsymbol{ \Gamma }_{kk'}^{\mathrm{real}} \big)
		\big( \mathrm{vec} \big( \boldsymbol{\Gamma}_{kk'}^{\mathrm{real}} \big) \big)^{\mathrm{H}} 
		\vphantom{\sum_{k=1}^{K}}
 		\nonumber
 		\\
		&
		\qquad\qquad\qquad
		+
		\mathrm{ vec } \big( \boldsymbol{ \Gamma }_{kk'}^{\mathrm{img}} \big)
		\big( \mathrm{vec} \big( \boldsymbol{ \Gamma }_{kk'}^{\mathrm{img}} \big) \big)^{\mathrm{H}}.
		\label{eq:GaBar}
\end{align}

Replacing the objective function \eqref{eq:OptWeCANObj} with \eqref{eq:WISLObjSubI}, the optimization problem \eqref{eq:OptWeCANI} can be rewritten as
\begin{subequations}
	\label{eq:OptWeCANII}
	\begin{alignat}{3}
		\underset{\mathbf{\tilde{Y}}} {\mathrm{min}}
		& \quad \big( \mathrm{vec} \big( \mathbf{\tilde{Y}} \big) \big)^{\mathrm{H}} \boldsymbol{ \tilde{\Phi}} \, \mathrm{vec} \big( \mathbf{\tilde{Y}}	\big) \label{eq:OptWeCanIIObj} \\
		\mathrm{s.t.}
		& \quad \mathbf{\tilde{Y} } = \mathbf{y} \mathbf{y}^{\mathrm{H}}, \quad \left| \mathbf{y} (p')  \right| = 1, \; p' = 1, \ldots, MP
	\end{alignat}
\end{subequations}
where \eqref{eq:OptWeCanIIObj} takes a quadratic form, to which a majorant can be applied. Yet before applying the majorization procedure, we present the following result that will be used later.

\begin{lemma} \label{LemmaII}
	Given a set of $N \times 1$ arbitrary complex vectors $\{ \mathbf{d}_k \}_{k=1}^{K}$ and an $N \times N$ arbitrary Hermitian matrix $\mathbf{H}$, the following generalized inequality
	\begin{align}
		\sum_{ k=1 }^{ K }	\left( \mathbf{d}_k	\mathbf{d}_k^{ \mathrm{ H } } \right)	\odot	\mathbf{H}	\preceq	\lambda_{ \mathrm{max} }	\left( \mathbf{H} \right)
		\mathbf{D}
		\label{eq:LemmaII}
	\end{align}
holds, where $\mathbf{D} \triangleq \mathrm{diag} \left \{ \sum_{ k=1 }^{ K }\left| \mathbf{d}_k \left( 1 \right) \right|^2, \ldots, \sum_{ k=1 }^{ K }\left| \mathbf{d}_k \left( N \right) \right|^2 \right \}$.
\end{lemma}
\begin{IEEEproof}
	Let $ \{ \tilde{\lambda}_n \}_{n=1}^{N} $ and $ \{ \mathbf{ \tilde{q} }_n \} _{n=1}^{N} $ be respectively the sets of eigenvalues (in descending order) and corresponding eigenvectors of the matrix $ \mathbf{H} $, i.e., $ \mathbf{H} = \sum_{n=1}^{N}	\tilde{ \lambda }_n	\tilde{\mathbf{q} }_n \tilde{\mathbf{q}}_n^{\mathrm{H}}$. Using this expression and elementary properties of the Hadamard product, the inequality \eqref{eq:LemmaII} can be derived as
	\begin{align}
		\sum_{ k=1 }^{ K }	\left( \mathbf{d}_k	\mathbf{d}_k^{ \mathrm{ H } } \right)	\odot	\mathbf{H}
		&
		=
		\left( \sum_{ k=1 }^{ K }	\mathbf{d}_k	\mathbf{d}_k^{ \mathrm{ H } } \right)	
		\odot	
		\left( \sum_{ n=1 }^{ N }	\tilde{ \lambda }_n		\tilde{ \mathbf{q} }_n	\tilde{ \mathbf{q} }_n^{ \mathrm{H} } \right)
		\nonumber
		\\
		&
		=
		\sum_{ k=1 }^{ K }	\sum_{ n=1 }^{ N }	
		\tilde{ \lambda }_n		\left( \mathbf{d}_k	\mathbf{d}_k^{ \mathrm{ H } } \right)	
		\odot
		\left( \tilde{ \mathbf{q} }_n	\tilde{ \mathbf{q} }_n^{ \mathrm{H} } \right)
		\nonumber
		\\
		&
		=
		\sum_{ k=1 }^{ K }	\sum_{ n=1 }^{ N }	
		\tilde{ \lambda }_n		\left( \mathbf{d}_k	\odot	\tilde{ \mathbf{q} }_n \right)	
		\left( \mathbf{d}_k \odot	\tilde{ \mathbf{q} }_n \right)^{ \mathrm{H} }
		\nonumber
		\\
		&
		\preceq
		\sum_{ k=1 }^{ K }	\sum_{ n=1 }^{ N }	
		\tilde{ \lambda }_1		\left( \mathbf{d}_k	\odot	\tilde{ \mathbf{q} }_n \right)	
		\left( \mathbf{d}_k \odot	\tilde{ \mathbf{q} }_n \right)^{ \mathrm{H} }
		\nonumber
		\\
		&
		=
		\tilde{ \lambda }_1		\left( \sum_{ k=1 }^{ K }	\mathbf{d}_k	\mathbf{d}_k^{ \mathrm{ H } } \right)	
		\odot
		\left( \sum_{ n=1 }^{ N } \tilde{ \mathbf{q} }_n	\tilde{ \mathbf{q} }_n^{ \mathrm{H} } \right)
		\nonumber
		\\
		&
		=
		\tilde{ \lambda }_1		\left( \sum_{ k=1 }^{ K }	\mathbf{d}_k	\mathbf{d}_k^{ \mathrm{ H } } \right)	
		\odot
		\mathbf{ I }_N
		\nonumber
		\\
		&
		=
		\lambda_{ \mathrm{max} }	\left( \mathbf{H} \right)  \mathbf{D}.
	\end{align}
	The proof is complete.
\end{IEEEproof}

Applying Lemma~\ref{LemmaII} by taking $\mathbf{d}_{k} = \mathrm{vec} \big( \mathbf{A}_{p} \mathbf{A }_{p}^{\mathrm{H}} \big)$, $\mathbf{H} = \boldsymbol{\bar{\Gamma}}$, and $K = 2P$, we obtain the following inequality 
\begin{align}
			\boldsymbol{ \bar{ \Phi }}	\odot	\boldsymbol{ \bar{ \Gamma } } 
			\preceq
			\lambda_{ \mathrm{ max } } \big( 	\boldsymbol{ \bar{ \Gamma } }  \big) 
			\mathrm{diag}	\left  \{ \boldsymbol{ \bar{ \Phi } }   \right  \}.
			\label{eq:IneqLemma}
\end{align}
Note that for a given matrix $\boldsymbol{\Gamma}$ in \eqref{eq:GaMatrix}, the largest eigenvalue of $\boldsymbol{\bar{\Gamma}}$ in \eqref{eq:GaBar}, i.e., $ \lambda_{\mathrm{max}} \big( 	\boldsymbol{\bar{\Gamma}} \big)$, is fixed, and it can be found that $ \lambda_{\mathrm{max}} \big( 	\boldsymbol{\bar{\Gamma}} \big) = \lambda_{\mathrm{max}}^{2} \big( 	\boldsymbol{ \Gamma} \big)  $. Moreover, the diagonal elements of $\boldsymbol{\bar{ \Phi}}$ take values either zero or $2P$. Therefore, we can replace the matrix $\mathrm{diag} \{ \boldsymbol{\bar{\Phi}} \}$ in \eqref{eq:IneqLemma} with an identity matrix magnified by $2P$ without disobeying the inequality. 

Using \eqref{eq:gxQua} with $\mathbf{G} \triangleq \lambda_{\boldsymbol{\tilde{\Phi}}} \mathbf{I}_{ M^2P^2}$ $\big( \text{here} \lambda_{\boldsymbol{\tilde{\Phi}}} \triangleq 2P \lambda_{\mathrm{max}} \big( \boldsymbol{\bar{\Gamma}} \big) \big)$ that satisfies $\mathbf{G} \succeq \boldsymbol{ \tilde{ \Phi}}$,
the objective function \eqref{eq:OptWeCanIIObj} can be majorized by the following function
\begin{align}
		\tilde{g}_1 
		&\big( \mathbf{\tilde{Y}}, \mathbf{\tilde{Y}}^{(k)} \big)
		 = \tfrac{\lambda_{\boldsymbol{\tilde{\Phi}}}}{2} \big( \mathrm{vec} \big( \mathbf{ \tilde{Y}} \big) \big)^{\mathrm{H}} \mathrm{vec} \big( \mathbf{\tilde{Y}}	\big) \nonumber \\
		& \quad + \big( \mathrm{vec} \big( \mathbf{\tilde{Y}}^{(k)} \big) \big)^{\mathrm{H}} \left( \tfrac{\lambda_{\boldsymbol{\tilde{\Phi}}}}{2} \mathbf{I}_{M^2P^2} - \boldsymbol{\tilde{ \Phi}} \right) \mathrm{vec} \big( \mathbf{\tilde{Y}}^{(k)} \big) \nonumber \\
		& \quad + 2 \Re \left\{ \big( \mathrm{vec} \big( \mathbf{\tilde{Y}} \big) \big)^{\mathrm{H}} \big( \boldsymbol{ \tilde{\Phi}} - \tfrac{\lambda_{\boldsymbol{\tilde{\Phi}}}}{2} \mathbf{I }_{M^2P^2} \big) \mathrm{vec} \big( \mathbf{ \tilde{Y}}^{(k)} \big) \right\} .
	\label{eq:MFWeCANI}
\end{align}
Due to the property \eqref{eq:ProVecConI}, the first and second terms in \eqref{eq:MFWeCANI} are constant and therefore immaterial for optimization. Ignoring these terms, \eqref{eq:OptWeCANII} can be majorized by the problem 
\begin{subequations}
	\label{eq:OptWeCANIII}
	\begin{alignat}{3}
		\underset{ \mathbf{\tilde{Y}}} {\mathrm{min}}
		& \quad \big( \mathrm{vec}^{\mathrm{H}} \big( \mathbf{\tilde{Y}} \big) \big)^{\mathrm{H}} \left( \boldsymbol{ \tilde{ \Phi } } - \tfrac{\lambda_{\boldsymbol{\tilde{\Phi}}}}{2} \mathbf{ I }_{ M^2 P^2 }	 \right)	\mathrm{ vec } \big( \mathbf{ \tilde{Y} }^{(k)} \big)
		\label{eq:OptWeCANIIIObj} \\
		\mathrm{ s.t. }
		& \quad \mathbf{\tilde{Y}} = \mathbf{y} \mathbf{y}^{\mathrm{H}}, \quad \left| \mathbf{y} (p')  \right| = 1, \; p' = 1, \ldots, MP.
	\end{alignat}
\end{subequations}

To further simplify \eqref{eq:OptWeCANIIIObj}, we will need the following result that relates Hadamard and Kronecker products.
\begin{lemma}	\label{LemmaIII}
	Given two matrices $\mathbf{F}$ and $\mathbf{C}$ of the same size $N \times N$ and the $N \times N^2$ selection matrix $\mathbf{E} = \big[ \mathbf{\bar{E}}_1, \ldots, \mathbf{\bar{E}}_{N} \big]$ with $\mathbf{\bar{E}}_n$ being the $ n $th $N \times N$ block matrix composed of all zeros except the $n$th element on the main diagonal equalling one, i.e., $[\mathbf{\bar{E}}_n]_{n,n} = 1$, the following equality
	\begin{align}
	\mathbf{ F } \odot	\mathbf{ C } = \mathbf{ E } \left( \mathbf{ C } \otimes \mathbf{ F }	\right) \mathbf{ E }^{ \mathrm{ H } } 
	\label{eq:EqlMatrPro}
	\end{align}
	holds. Under the condition that $ \sqrt{N} $ is an integer, $\mathbf{\bar{E}}_n$ can be decomposed as 
\begin{align}
	\mathbf{ \bar{ E } }_n = \mathbf{ \hat{ E } }_{u\left( n \right)} \otimes \mathbf{ \hat{E} }_{v \left( n \right) } 
	\label{eq:SelMatrxM}
\end{align}
where the matrices $\mathbf{\hat{E}}_{u(n)} $ and $\mathbf{\hat{E}}_{v(n)}$ are constructed in the same way as $\mathbf{\bar{E}}_{n}$ but have the reduced size $\sqrt{N} \times \sqrt{N}$, and
\begin{align}
	u \left( n \right )
	&
	\triangleq
	\left \lfloor \frac{n-1}{\sqrt{N}}	\right \rfloor	+ 1, \; n = 1, \ldots, N
	\\
	v \left (  n  \right  )
	&
	\triangleq
	\mathrm{mod} \left ( n-1, \sqrt{N}  \right ) + 1, \; n = 1, \ldots, N
	\label{eq:SelMatrxEnd}
\end{align}
are respectively the column and row indices of the element in the $ \sqrt{N} \times \sqrt{N}  $ matrix with linear (column-wise) index $n$. 
	\begin{IEEEproof}
		The proof of \eqref{eq:EqlMatrPro} appears in Lemma~1 of \cite{LiuMatrixPro08}. The remaining results \eqref{eq:SelMatrxM}--\eqref{eq:SelMatrxEnd} are the elementary properties of the selection matrix.
	\end{IEEEproof}
\end{lemma}

Applying Lemma~\ref{LemmaIII} by taking $\mathbf{F} = \mathbf{\bar{\Phi}}$, $\mathbf{C} = \mathbf{ \bar{\boldsymbol{\Gamma}}}$, and $N = M^2 P^2$, and substituting \eqref{eq:PhiTilDef} into \eqref{eq:OptWeCANIIIObj}, the objective function \eqref{eq:OptWeCANIIIObj}, denoted for brevity as $ \mathrm{obj}^{\mathrm{d}} $, can be rewritten as
\begin{align}
	\mathrm{obj}^{\mathrm{d}} 
	& = 
	\big( \mathrm{vec} \big( \mathbf{\tilde{Y}} \big) \big)^{\mathrm{H}} \left( \mathbf{E} \left( \boldsymbol{\bar{\Gamma}} \otimes \boldsymbol{\bar{\Phi}} \right)\mathbf{E}^{\mathrm{H}} - \tfrac{ \lambda_{\boldsymbol{\tilde{\Phi}}}}{2} \mathbf{I}_{M^2 P^2}	\right) \mathrm{vec} \big( \mathbf{ \tilde{Y}}^{(k)} \big) 
	\nonumber\\
	& = \big( \mathrm{vec} \big( \mathbf{\tilde{Y}} \big) \big)^{\mathrm{H}}	\left( \sum_{n=1}^{M^2 P^2} \sum_{n'=1}^{M^2 P^2} \left[ \boldsymbol{\bar{\Gamma}} \right]_{n, n'} \mathbf{ \bar{E} }_n \boldsymbol{\bar{ \Phi}} \mathbf{ \bar{E}}_{n'}^{\mathrm{H}} \right) \mathrm{ vec } \big( \mathbf{ \tilde{Y}}^{(k)} \big) \nonumber \\
	& \quad\; - \tfrac{\lambda_{\boldsymbol{\tilde{\Phi}}}}{2} \big( \mathrm{vec} \big( \mathbf{\tilde{Y}} \big) \big)^{\mathrm{H}}	\mathrm{vec} \big( \mathbf{ \tilde{Y}}^{(k)} \big) \label{eq:LongEqII}
\end{align}
where the latter expression in \eqref{eq:LongEqII} is obtained by expanding the Kronecker product in the prior expression for the objective.
	
Using \eqref{eq:GaBar} and \eqref{eq:SelMatrxM}, and applying the properties \eqref{eq:PropI} and \eqref{eq:ProII}, the objective \eqref{eq:LongEqII} can be further rewritten as
	\begin{align}
	 \mathrm{obj}^{\mathrm{d}} = &
			\sum_{ p=1 }^{ 2P }	\sum_{ n=1 }^{ M^2 P^2 }	\sum_{ n'=1 }^{ M^2 P^2 }	 
			  \left[ \boldsymbol{ \bar{ \Gamma } } \right]_{ n, n' } 	
			\mathbf{ y  }^{ \mathrm{H} } 	\big( \mathbf{ y }^{ \mathrm{ T } } 	\otimes  \mathbf{ I }_{ M P } \big)		\big( \mathbf{ \hat{E} }_{u\left( n \right)} 	
			\nonumber
			\\
			& 
				\otimes \mathbf{ \hat{E} }_{v \left( n \right) } \big)
				\big( \mathbf{ A }_p^{ \mathrm{ T } } \otimes \mathbf{ I }_{MP} \big)^{ \mathrm{ H } }		\mathrm{vec} \big( \mathbf{ A }_p  \big) \left( \mathrm{vec} \big( \mathbf{ A }_p \big) \right)^{\mathrm{H}} \big( \mathbf{ A }_p^{ \mathrm{ T } } 
			\nonumber
			\\
			& 
			\otimes \mathbf{ I }_{MP} \big)	
			\big( \mathbf{ \hat{E} }_{u\left( n' \right)} 	\otimes \mathbf{ \hat{E} }_{v \left( n' \right) } \big)^{ \mathrm{ H } }	
			\big( ( \mathbf{ y }^{(k)} )^{ \mathrm{ T } } 	\otimes  \mathbf{ I }_{ M P } \big)^{ \mathrm{ H } }		\mathbf{ y }^{(k)}
			\nonumber
			\\
			& - \tfrac{\lambda_{\boldsymbol{\tilde{\Phi}}}}{2} \mathbf{y}^{\mathrm H } 
			\big( \mathbf{y}^{ \mathrm{ T } }  \otimes  \mathbf{I}_{MP} \big)
			\big( ( \mathbf{y}^{(k)} )^{ \mathrm{ T } }  \otimes  \mathbf{I}_{MP} \big)^{ \mathrm{ H } }	\mathbf{y}^{(k)}.
			\label{eq:LongEqIII}
\end{align}
Applying the mixed-product property of the Kronecker product together with the property $\big( \big( \mathbf{y}^{\mathrm{T}}	\mathbf{\hat{E}}_{u(n)} \mathbf{A}_{p}^{\ast} \big) \otimes \mathbf{\hat{E} }_{v(n)} \big) \mathrm{vec} \big( \mathbf{A}_{p}\big ) = \mathbf{ \hat{E} }_{v(n)} \mathbf{A}_{p}	\mathbf{A}_{p}^{\mathrm{H}} \mathbf{\hat{E}}_{u(n)} \mathbf{y}$ to \eqref{eq:LongEqIII}, we obtain
	\begin{align}
		\mathrm{obj}^{\mathrm{d}}
			&
			=
				\mathbf{ y  }^{ \mathrm{H} } 
				\left(
				\sum_{ p=1 }^{ 2P } 
				\sum_{ n =1 }^{ M^2 P^2 }	\sum_{ n'=1 }^{ M^2 P^2 }	 
				\left[ \boldsymbol{ \bar{ \Gamma } } \right]_{ n, n' } 	
				\mathbf{ \hat{E} }_{v\left( n \right)} 	\mathbf{ A }_p		
				\left(
				\big( \mathbf{ y }^{(k)} \big)^{ \mathrm{ H } }
				\mathbf{ \hat{E} }_{u\left( n' \right)} 			
				\right.
				\right.
				\nonumber
				\\
				&
				\quad
				\times
				\!\! 
				\left. \vphantom{\sum_{l=1}^{M^2 P^2}}
				\left.
				\mathbf{ A }_p \mathbf{ A }_p^{ \mathrm{ H } }	
				\mathbf{ \hat{E} }_{v\left( n' \right)} 	\mathbf{ y }^{(k)} 
				\right)
				\mathbf{ A }_p^{ \mathrm{ H } }		\mathbf{ \hat{E} }_{u\left( n \right)} 	
				\!
				- \tfrac{\lambda_{\boldsymbol{ \tilde{ \Phi } } }}{2}
				\mathbf{ y  }^{ (k) }  ( \mathbf{ y  }^{ (k) } )^{ \mathrm{ H } }  \big)
				\!\!
				\right)
				\mathbf{ y  } \nonumber 
				\\
				&
				=
				\mathbf{ y  }^{ \mathrm{H} } 
				\left (		\mathbf{B}^{(k)} 
				- \tfrac{\lambda_{\boldsymbol{ \tilde{ \Phi } } }}{2} \mathbf{ y  }^{ (k) }  ( \mathbf{ y  }^{ (k) } )^{ \mathrm{ H } }
					\right )
				\mathbf{ y  }	
				\label{eq:OptWeCANIVObj}
\end{align}
where $\mathbf{y}^{(k)} \triangleq \mathrm{vec} \{ \mathbf{Y}^{(k)} \}$
and $\mathbf{B}^{(k)}$ is an $MP \times MP$ Hermitian matrix composed of $ M^2 $ block matrices, i.e.,
\begin{align}
	\mathbf{B}^{(k)}
  \triangleq
  \left[\!\!\!
  \begin{array}{ccc}
      \mathbf{B}^{(k)}_{11}    &     \ldots              &     \mathbf{B}^{(k)}_{1M} \\
      \vdots                         &     \ddots             &      \vdots          \\
      \mathbf{B}^{(k)}_{M1}            &      \ldots     &       \mathbf{B}^{(k)}_{MM}
  \end{array}
  \!\!\!\right]
  \label{eq:BkMatrix}
\end{align}
with the $ (m, m') $th block 
\begin{align}
	\mathbf{B}_{m m'}^{(k)}
	=
	2 P
	\mathcal{T} \left (	\boldsymbol{\rho}_{m m'}^{(k)}, \boldsymbol{\eta}_{m m'}^{(k)}	\right )
\end{align}
being a $ P \times P $ Toeplitz matrix whose first row and column coincide with the $P \times 1$ vectors $\boldsymbol{\rho}_{m m'}^{(k)}$ and $\boldsymbol{\eta}_{m m'}^{(k)}$, respectively. Here, the $(p+1) $th ($ 0 \leq p \leq P-1 $) elements of $ \boldsymbol{\rho}_{m m'}^{(k)} $ and $ \boldsymbol{\eta}_{m m'}^{(k)} $ are respectively given by
\begin{align} 
		\boldsymbol{\rho}_{m m'}^{(k)}
		(p+1)
		&
		\triangleq
		\begin{dcases*}
				\gamma_{p}^{2}
				\mathbf{1}_{P-p}^{ \mathrm{ T } }
				\mathcal{U}_{p} \! \left(	 \mathbf{Z}_{m m'}^{(k)} \right)	
				,	 & $ p \in \boldsymbol{\Omega} $
		  	 \\
		  	0,
		  	\vphantom{\mathrm{diag}^{ \mathrm{ T } } \left\{\right\}	}
		  	&
		  	$ p \in \boldsymbol{\bar{\Omega}} $
			\end{dcases*}
			\label{eq:rho}
			\\
					\boldsymbol{\eta}_{m m'}^{(k)}
					(p+1)
					&
					\triangleq
					\begin{dcases*}
							\gamma_{p}^{2}
							\mathbf{1}_{P-p}^{ \mathrm{ T } }
							\mathcal{D}_{p} \! \left(	 \mathbf{Z}_{m m'}^{(k)}	 \right)	
							,	 & $ p \in \boldsymbol{\Omega} $
					  	 \\
					  	0,
					  	\vphantom{\mathrm{diag}^{ \mathrm{ T } } \left\{			 \right\}	}
					  	&
					  	$ p \in \boldsymbol{\bar{\Omega}} $
						\end{dcases*}
						\label{eq:eta}
\end{align}
where $\mathbf{Z}_{m m'}^{(k)} \triangleq \mathbf{y}_{m}^{(k)} \big( \mathbf{y}_{m'}^{(k)} \big)^{ \mathrm{H}}$, $\boldsymbol{\Omega} \triangleq \{ 0 \} \cup \{ p | \gamma_{p} \neq 0, p > 0 \}$ is the set of non-negative indices associated with the non-zero ISL controlling weights (always including zero index for the sake of simplicity), and $\boldsymbol{\bar{\Omega}} \triangleq \{ p | \gamma_{p} = 0, p > 0 \}$ is the complementary set of $\boldsymbol{\Omega}$ with the full set defined as $[0, P-1]$. The meanings of \eqref{eq:rho} and \eqref{eq:eta} are that the non-zero elements of $ \boldsymbol{\rho}_{m m'}^{(k)} $ and $ \boldsymbol{\eta}_{m, m'}^{(k)} $ are expressed by the sum of the off diagonal elements in the upper and lower triangular parts of $\mathbf{Z}_{m m'}^{(k)}$ magnified by $\gamma_{p }^{2} $, respectively. Using \eqref{eq:rho} and \eqref{eq:eta}, we can avoid calculations for the zero elements. Note that $\mathbf{B}_{m m'}^{(k)} = \big (	\mathbf{B}_{ m' m }^{(k)}	\big)^{\mathrm{H}}$, therefore, only the upper (or lower) triangular part of $\mathbf{B}^{(k )}$ needs to be determined.

The objective function \eqref{eq:OptWeCANIVObj} takes a quadratic form, to which the majorant of \eqref{eq:gxQua} can be applied again. Let $\mathbf{G} \triangleq \tau^{(k)} \mathbf{I}_{MP}$, so that the generalized inequality $\mathbf{G} \succeq \mathbf{Q} $ is guaranteed for \eqref{eq:gxQua}. Here we can use any matrix norm of $\mathbf{Q}^{(k)} \triangleq \mathbf{B}^{(k)} - \lambda_{\boldsymbol{ \tilde{\Phi}}} \mathbf{y}^{(k)} ( \mathbf{y}^{(k)} )^{\mathrm{H}} / 2$ for $\tau^{(k)}$ because any matrix norm serves as an upper bound of the largest eigenvalue. Thus, the objective function \eqref{eq:OptWeCANIVObj} can be majorized by the following function
\begin{align}
		\tilde{g}_2  \big( \mathbf{ y }, & \mathbf{ y }^{(k)} \big)
		 \!=\! \tfrac{\tau^{(k)}}{2} \mathbf{y}^{ \mathrm{H}} \mathbf{y} \!+\! ( \mathbf{y}^{(k)} )^{\mathrm{ H}} \left ( \! \tfrac{\tau^{(k)} + M P \lambda_{ \boldsymbol{ \tilde{\Phi}}}}{2} \mathbf{I}_{MP}	\!-\! \mathbf{B}^{(k)} \! \right) \mathbf{y}^{(k)} \nonumber \\
		& \;\;\;\; + 2 \Re \left\{ \mathbf{y}^{ \mathrm{H}} \left( \mathbf{B}^{(k)} - \tfrac{\tau^{(k)} + M P \lambda_{ \boldsymbol{ \tilde{ \Phi}}}}{2} \mathbf{I}_{MP} \right) \mathbf{y}^{(k)} \right\}. \label{eq:MFWeCANIINew}
\end{align}
Similar to the majorant \eqref{eq:MFCANII}, the first two terms of \eqref{eq:MFWeCANIINew} are constant and therefore immaterial for optimization. Ignoring these two terms, the problem \eqref{eq:OptWeCANIII} can be majorized by
\begin{align}
  \underset{ \mathbf{ y } } {  \mathrm{ min } }
  & \quad \mathbf{y}^{ \mathrm{ H } } \left( \mathbf{B}^{(k)} - \tfrac{\tau^{(k)} + M P \lambda_{ \boldsymbol{ \tilde{\Phi}}}}{2} \mathbf{I}_{MP} \right) \mathbf{y}^{(k)} \nonumber \\
  \mathrm{ s.t. }
  & \quad \left| \mathbf{ y }   (p')  \right| = 1, \; p' = 1, \ldots, MP. \label{eq:OptWeCANV}
\end{align}
Due to the constant modulus property of $\mathbf{y}$, the problem \eqref{eq:OptWeCANV} is equivalent to the following optimization problem
\begin{align}
	\underset{ \mathbf{ y } } {  \mathrm{ min } }
	& \quad \big\| \mathbf{ y } - \mathbf{ z }^{ (k) } \big\|^2 \nonumber \\
	\mathrm{ s.t. }
	& \quad \left| \mathbf{ y } (p')  \right| = 1, \; p' = 1, \ldots, MP. \label{eq:OptWeCANVI}
\end{align}
where $\mathbf{ z }^{ (k) } \triangleq \left( \big( \tau^{(k)} + M P \lambda_{ \boldsymbol{ \tilde{ \Phi } } } \big) \mathbf{I}_{MP} / 2 - \mathbf{B}^{(k)} \right) \mathbf{y}^{(k)}$. The problem \eqref{eq:OptWeCANVI} can be then solved in closed form as 
\begin{align}
  \mathbf{y} \big( p' \big) = \exp \{ j \cdot \mathrm{arg} \big( \mathbf{z}^{(k)} (p') \big) \}, \; p' = 1, \ldots, MP.
  \label{eq:SolWeCAN}
\end{align}
Finally, reshaping the so-obtained vector $\mathbf{y}$ into a $P \times M$ matrix, we obtain the designed waveform matrix $\mathbf{Y}$. The WISL minimization based unimodular waveform design algorithm is summarized in Algorithm~\ref{WeCANMM}.

\begin{figure}[!t]%
	\vspace{-7pt}
	\begin{algorithm}[H]
		\caption{The WISL Minimization-Based Algorithm}
		\label{WeCANMM}
		\begin{algorithmic}[1]
			\myState {$k \leftarrow 0$, $ \mathbf{ y }\leftarrow$ unimodular sequence with random phases.}
			\myState{$ \lambda_{\boldsymbol{ \tilde{ \Phi } } } \triangleq  2 M P \lambda_{ \mathrm{ max } }^{2} \big( 	\boldsymbol{ \Gamma }  \big)  $}
			\Repeat
			\Procedure {WISLMaMi}{}{$\big(\mathbf{ y }^{ ( k )  } \big)$}
			\myState{$\begin{aligned}
				&\text{Calculate  }\boldsymbol{\rho}_{m m'}^{(k)} ,  \,  \boldsymbol{\eta}_{m m'}^{(k)} \text{ via }\eqref{eq:rho}\text{ and } \eqref{eq:eta},\\
				& \qquad\qquad\qquad\quad      m = 1, \ldots, M; m' = m, \ldots, M. 
				\end{aligned}
				$}
			\myState{$ \begin{aligned}
				&\mathbf{B}_{m m'}^{(k)}
				=
				\left ( 	\mathbf{B}_{m' m}^{(k)}	\right )^{ \mathrm{ H } }
				=
				2 P
				\mathcal{T} \left (	\boldsymbol{\rho}_{m m'}^{(k)}, \boldsymbol{\eta}_{m m'}^{(k)}	\right ), \\
				& \qquad\qquad\qquad\quad      m = 1, \ldots, M; m' = m, \ldots, M. 
				\end{aligned}
				$}
			\myState{Construct $ \mathbf{B}^{(k)} $ via \eqref{eq:BkMatrix}}
			\myState{$ 
				\tau^{(k)}
				=
				\left \|
				\mathbf{B}^{(k)} 
				-
				\tfrac{1}{2}
				\lambda_{\boldsymbol{ \tilde{ \Phi } } } 	  \mathbf{ y  }^{ (k) }  ( \mathbf{ y  }^{ (k) } )^{ \mathrm{ H } }
				\right \|
				$}
			\myState{$
				\begin{aligned}
				\mathbf{ z }^{ (k) }   
				\!
				&
				=
				\left (
				\tfrac{1}{2}
				\big( 	\tau^{(k)}  + M P 	\lambda_{ \boldsymbol{ \tilde{ \Phi } } }  \big)   \mathbf{I}_{MP}
				-
				\mathbf{B}^{(k)}
				\right  )
				\mathbf{y}^{(k)}
				\end{aligned}$}
			\myState{$  \mathbf{ y }^{ (k+1) } (p') = e^{ j  \mathrm{ arg } \left(  \mathbf{ z }^{ (k) } (p') \right)  }, \; p' = 1, \ldots, MP $}
			\myState{$ k \leftarrow k+1$}
			\EndProcedure
			\Until convergence
		\end{algorithmic}
	\end{algorithm}
	\vspace{-20pt}
\end{figure}

\begin{figure*}[t]
\centering
\subfloat[Normalized ISL values versus the number of conducted iterations. The stopping criterion (i) is used.\label{Fig:ISLVSITE_a}] { \includegraphics[width=0.464\textwidth]{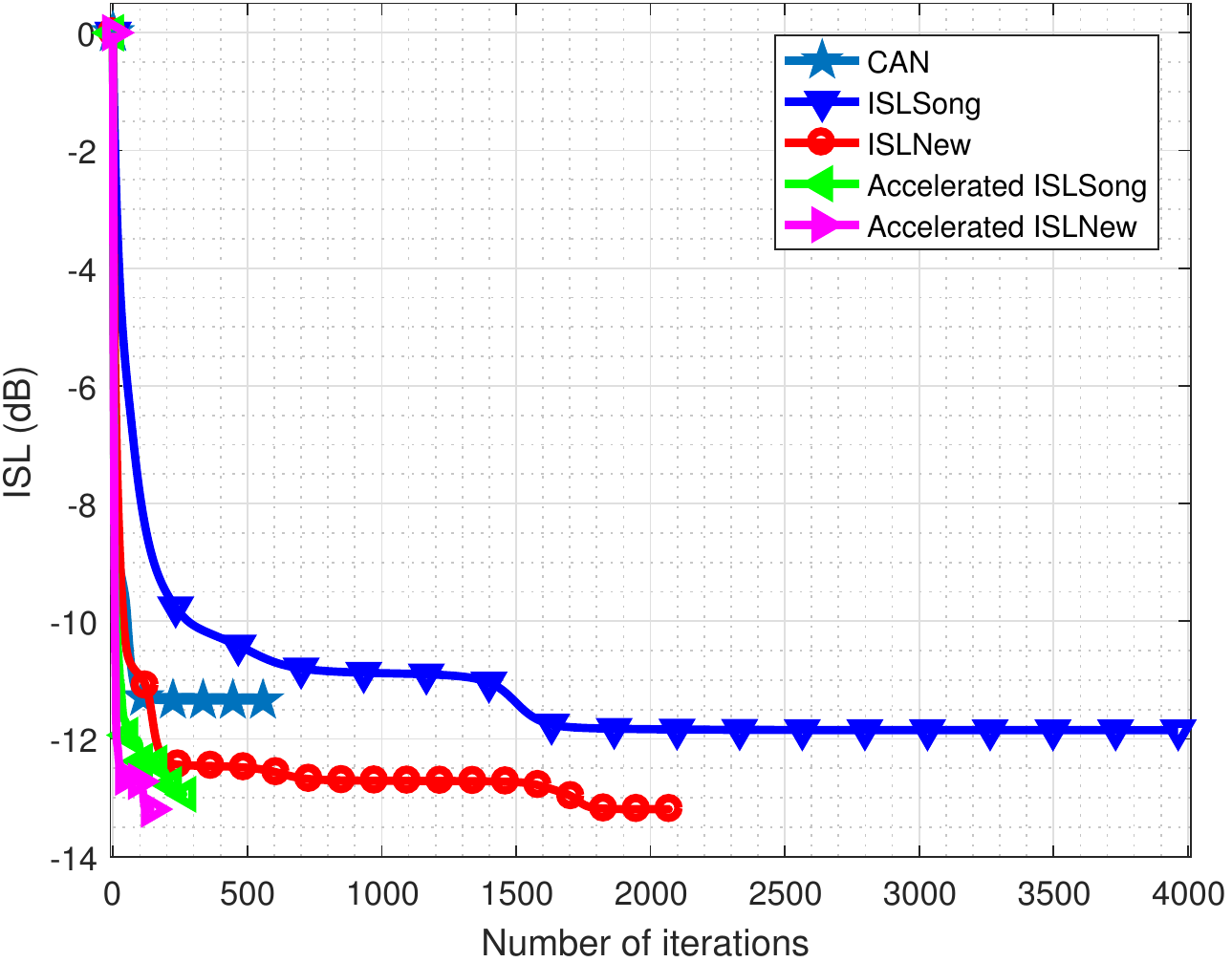} }
\hfill
\subfloat[Normalized ISL values versus the number of conducted iterations. The stopping criterion (ii) is used.\label{Fig:ISLVSITE_b}] { \includegraphics[width=0.45\textwidth]{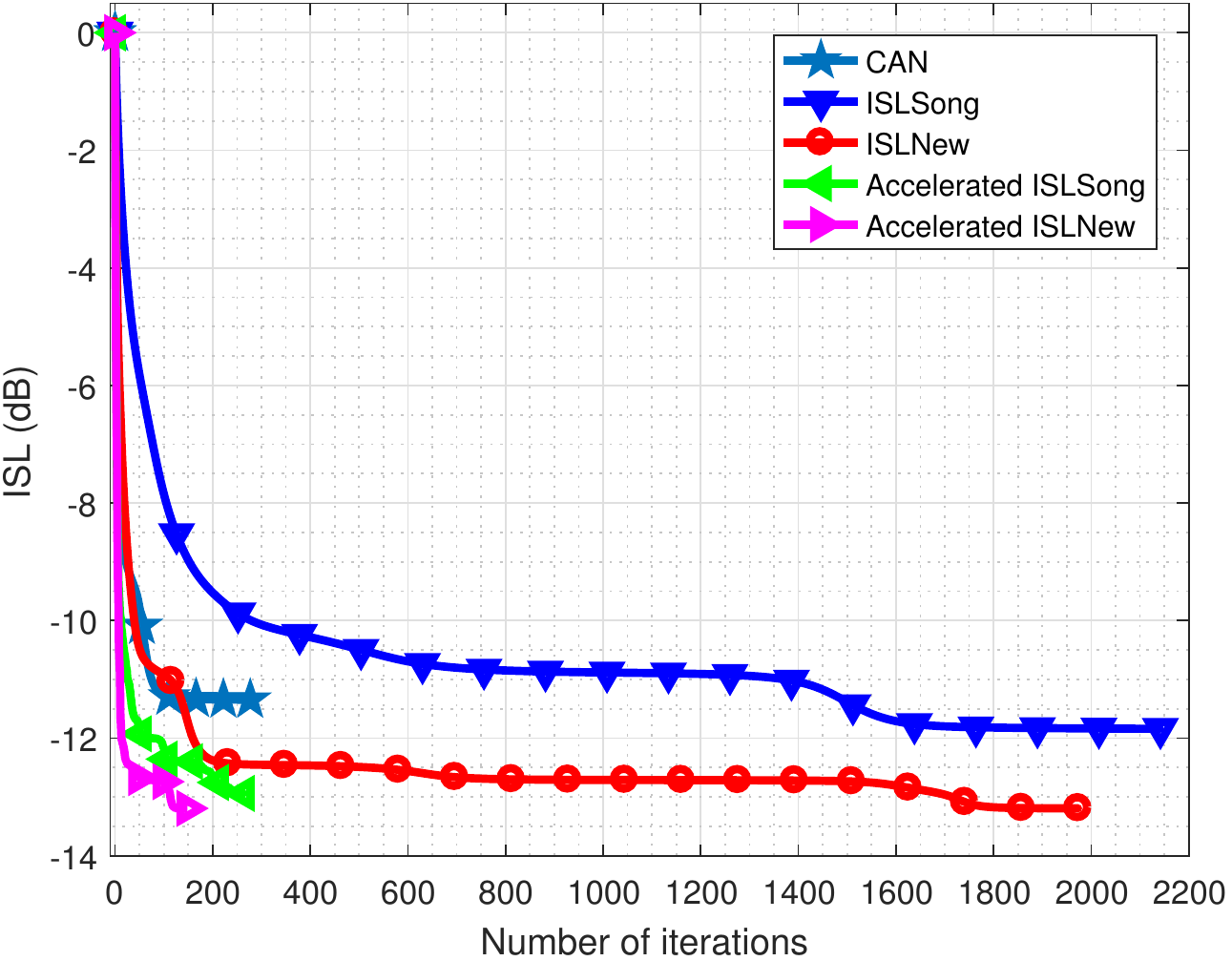} }
\caption{Convergence of the ISL minimization-based algorithms. First simulation example: $M = 1$ and $P = 128$.}
\label{Fig:ISLVSITE}
\vspace{-4pt}
\end{figure*}

To find the computational complexity of Algorithm~\ref{WeCANMM}, we assume that the set of $\boldsymbol{\Omega}$ consists of $N_{P} \, (0<N_{P}\leq P)$ elements. It can be seen that both $\boldsymbol{\rho}_{m m'}^{(k)}$ in \eqref{eq:rho} and $ \boldsymbol{\eta}_{m m'}^{(k)} $ in \eqref{eq:eta} can be calculated with at most $N_{P} P$ operations if $\mathbf{Z}_{m m'}^{(k)}$ is given. The calculation of the covariance matrix $\mathbf{Z}_{m m'}^{(k)}$ costs $P^{2}$ operations. Note that we only need to do calculations for the subscripts $m = 1, \ldots, M$ and $m' = m, \ldots, M$, and then repeat the above summarized calculations $M(M-1)/2$ times. Finally, the calculation of the vector $\mathbf{z}^{(k)}$ needs $M^2 P^2$ operations. Consequently, the total number of operations is upper bounded by $((3M^2 - M) P^2  + (M^2-M) N_P P) / 2$. In other words, the computational complexity of Algorithm~\ref{WeCANMM} is at most $\mathcal{O}((M-1)MP^2)$ that is smaller than quadratic in the problem size, and therefore suitable for large-scale optimization.\footnote{Note that the overall computational complexity claimed in \cite{SongWFMaMi16} has to be corrected to the same as here. Further clarifications can be found in Section~\ref{Sec:Simulation}.} 

The accelerated version of Algorithm~\ref{WeCANMM} is obtained by a straightforward application of the SQUAREM acceleration scheme \cite{Varadhan08} as in the case of Algorithm~\ref{CANMM}.

\section{Simulation Results}\label{Sec:Simulation}
We evaluate here the performance of the proposed ISL and WISL minimization-based waveform design algorithms (Algorithms~\ref{CANMM}~and~\ref{WeCANMM}) by comparing them with existing ISL and WISL minimization based algorithms. To be specific, our Algorithm~\ref{CANMM} for ISL minimization (named hereafter as ISLNew) is compared with the CAN of \cite{HeMIMO09} and the (third) algorithm in \cite{SongWFMaMi16} (named hereafter as ISLSong), while our Algorithm~\ref{WeCANMM} for WISL minimization (named hereafter as WISLNew) is compared with the WeCAN of \cite{HeMIMO09} and the (second) algorithm in \cite{SongWFMaMi16} (named hereafter as WISLSong). The accelerated versions of the MaMi-based algorithms, including the ISLNew, WISLNew, ISLSong, and WISLSong algorithms, are also tested, where the SQUAREM scheme \cite{Varadhan08} is used for MaMi acceleration. We generate sets of unimodular sequences with random phases as the initialization for each algorithm tested, and apply the same set of sequences to all algorithms for the purpose of fair comparison. All comparisons are conducted based on the same hardware and software platforms. Throughout our simulations, two stopping criteria are employed: (i) the absolute ISL or WISL difference between the current and previous iterations normalized by the initial ISL or WISL, whose tolerance is set to be $ 10^{-8} $; and (ii) the norm  difference between the waveform matrices (or vectors) obtained at the current and previous iterations, whose tolerance is set to be $ 10^{-3} $. The ISL and WISL values in dBs are defined as $ 10\log_{10}(\zeta) $ and $ 10\log_{10}(\zeta_{\mathrm{w}}) $, respectively.

\subsection{ISL Minimization}
In the first example, we study the convergence properties of the waveform design algorithms (CAN, ISLSong, ISLNew, accelerated ISLSong, and accelerated ISLNew) in terms of the number of conducted iterations for a problem of relatively small size. Specifically, a single waveform ($ M=1 $) of the code length $ P = 128 $ is designed in this example. In  Figs.~\ref{Fig:ISLVSITE}\subref{Fig:ISLVSITE_a} and \ref{Fig:ISLVSITE}\subref{Fig:ISLVSITE_b}, the ISL performance versus the number of conducted iterations is displayed for the aforementioned algorithms, where the stopping criteria (i) and (ii) are used, respectively. The ISL values for different algorithms obtained at each iteration are normalized by the ISL value associated with the initial set of sequences. 

It can be seen from Figs.~\ref{Fig:ISLVSITE}\subref{Fig:ISLVSITE_a} and \ref{Fig:ISLVSITE}\subref{Fig:ISLVSITE_b} that for all the algorithms tested, the ISL decreases monotonically as the number of iterations increases. Among the ISL minimization-based algorithms tested, the accelerated ISLNew algorithm shows the best convergence speed, i.e., it requires the smallest number of iterations to converge to a solution that satisfies the pre-set tolerance parameter for both stopping criteria used. The accelerated ISLSong algorithm shows the second best convergence speed. This demonstrates the superiority of applying accelerated MaMi techniques to the ISL minimization-based waveform design. The proposed ISLNew algorithm without acceleration shows a little slower convergence speed, but achieves around $ 4 $~dB better ISL than that of the CAN algorithm. The ISLSong algorithm without acceleration shows the worst convergence speed among all the algorithms tested. The same convergence behavior can also be seen in Figs.~\ref{Fig:ISLVSITE}\subref{Fig:ISLVSITE_a} and  \ref{Fig:ISLVSITE}\subref{Fig:ISLVSITE_b} independent on the stopping criteria used. 
 
\begin{table*}[!th]
	\centering
	\begin{adjustbox}{max width=\textwidth}
		\begin{threeparttable}
			\caption{ISL performance comparisons (including the excess minimum and average ISL, the consumed time, and the number of conducted iterations) of the algorithms tested versus code length for $ M=2 $ waveforms for stopping criterion (i)}
			\abovecaptionskip = 7pt
			\label{Tab:ISL_T1}
			\centering
			\setlength\tabcolsep{2pt}
			\begin{tabular}{|c|c|c|c|c|c|c|c|c|c|c|c|c|c|c|c|c|c|c|c|c|c}
				\hline
				\multicolumn{1}{|c|}{\multirow{2}{*}{}} 	& \multicolumn{4}{c|} {$ P=32, M=2 $} 	& \multicolumn{4}{c|} {$ P=128, M=2 $}	& \multicolumn{4}{c|} {$ P=512, M=2 $}	& \multicolumn{4}{c|} {$ P=1024, M=2 $} 		& \multicolumn{4}{c|} {$ P=2048, M=2 $}\\ \cline{2-21}
				&  Min.\tnote{\emph{a}}\,  & Ave.\tnote{\emph{b}}\,                       &  Time                      & Iter.\tnote{\emph{c}}\, &  Min.  & Ave.                        &  Time                      & Iter. &  Min.  & Ave.                        &  Time                      & Iter. &  Min.  & Ave.                        &  Time                      & Iter. &  Min.  & Ave.                        &  Time                      & Iter. \\	\hline
				CAN			&	33.11		&	33.11	&	1.17		& 5141	&	45.15		&	45.16	&	5.99		& 8002	&	57.20		&	57.20	&	18.34		& 6975 &	63.22		&	63.22	&	33.56		& 6729  &	 69.24		&	69.24	&	72.99		& 6990 \\ \hline
				ISLSong			&	33.11		&	33.11	&	0.11		& 438	&	45.16		&	45.16	&	0.50		& 489 &	57.20		&	57.20	&	5.01		& 491	&	63.22		&	63.22	&	19.11		& 461		&	69.24	&	69.24	&	62.47		& 491	\\ \hline
				ISLNew			&	33.11	&	33.11	&	0.07		& 375	&	45.16		&	45.16	&	0.34		& 410 &	57.20		&	57.20	&	4.30	& 456	&	63.22		&	63.22	&	18.28		& 444	&	69.24		&	69.24	&	55.14		& 469	\\ \hline
			\end{tabular}
			\begin{tablenotes}
				\item[\emph{a}]Min.: Obtained minimum ISL value (in dB).
				\item[\emph{b}]Ave.: Obtained average ISL value (in dB).
				\item[\emph{c}]Iter.: Number of conducted iterations.
			\end{tablenotes}
		\end{threeparttable}
	\end{adjustbox}
\end{table*}  

\begin{table*}[!th]
	\centering
	\begin{adjustbox}{max width=\textwidth}
		\begin{threeparttable}
			\caption{ISL performance comparisons (including the excess minimum and average ISL, the consumed time, and the number of conducted iterations) of the algorithms tested versus code length for $ M=2 $ waveforms for stopping criterion (ii)}
			\abovecaptionskip = 7pt
			\label{Tab:ISL_T2}
			\centering
			\setlength\tabcolsep{2pt}
			\begin{tabular}{|c|c|c|c|c|c|c|c|c|c|c|c|c|c|c|c|c|c|c|c|c|c}
				\hline
				\multicolumn{1}{|c|}{\multirow{2}{*}{}} 	& \multicolumn{4}{c|} {$ P=32, M=2 $} 	& \multicolumn{4}{c|} {$ P=128, M=2 $}	& \multicolumn{4}{c|} {$ P=512, M=2 $}	& \multicolumn{4}{c|} {$ P=1024, M=2 $} 		& \multicolumn{4}{c|} {$ P=2048, M=2 $}\\ \cline{2-21}
				&  Min.\tnote{\emph{a}}\,  & Ave.\tnote{\emph{b}}\,                       &  Time                      & Iter.\tnote{\emph{c}}\, &  Min.  & Ave.                        &  Time                      & Iter. &  Min.  & Ave.                        &  Time                      & Iter. &  Min.  & Ave.                        &  Time                      & Iter. &  Min.  & Ave.                        &  Time                      & Iter. \\	\hline
				CAN			&	33.11		&	33.12	&	0.35		& 1605	&	45.15		&	45.16	&	 3664		& 8002	&	57.20		&	57.20	&	19.74		& 7863 &	63.22		&	63.22	&	54.53		& 11272  &	 69.24		&	69.24	&	163.89		& 16846 \\ \hline
				ISLSong			&	33.11		&	33.11	&	0.11		& 401	&	45.15		&	45.15	&	0.11		& 896 &	57.20		&	57.20	&	17.22		& 1788	&	63.22		&	63.22	&	114.13		&  2758		&	69.24	&	69.24	&	633.45		& 3802	\\ \hline
				ISLNew			&	33.11	&	33.11	&	0.05		& 223	&	45.15		&	45.15	&	0.43		& 527 &	57.20		&	57.20	&	8.25	& 938	&	63.22		&	63.22	&	54.49		& 1338	&	69.24		&	69.24	&	163.89		& 1798	\\ \hline
			\end{tabular}
			\begin{tablenotes}
				\item[\emph{a}]Min.: Obtained minimum ISL value (in dB).
				\item[\emph{b}]Ave.: Obtained average ISL value (in dB).
				\item[\emph{c}]Iter.: Number of conducted iterations.
			\end{tablenotes}
		\end{threeparttable}
	\end{adjustbox}
	\vspace{-10pt}
\end{table*}
   
In the second example, we use another way of evaluating the performance of the algorithms tested by comparing them with each other in term of the following performance characteristics: the minimum and average ISL values obtained after the corresponding algorithms converge to satisfy the pre-set tolerance, called hereafter as ISL after convergence (in dBs); the average consumed time (in seconds); and the average number of conducted iterations. Different from the previous example, where the results have been shown for a random trial, the results are averaged over $ 50 $ independent trials in this example. The number of designed waveforms is fixed to $ M = 2 $ and the code length $ P $ takes values from the set $ \{32, 128, 512, 1024, 2048\} $. The results for the CAN, accelerated ISLSong, and accelerated ISLNew algorithms are shown in Tables~\ref{Tab:ISL_T1}~and~~\ref{Tab:ISL_T2} for, respectively, the stopping criteria (i) and (ii).

It can be seen from Table~\ref{Tab:ISL_T1} that the CAN, accelerated ISLSong, and accelerated ISLNew algorithms obtain nearly the same minimum and average ISL values after convergence for all code lengths, which indicates that the latter two algorithms show no significant advantages over the benchmark CAN algorithm in terms of the ISL performance. This is mainly because all the three algorithms converge near to the lower bound of the minimum achievable ISL for each code length. Thus, there is little room left for improvement. The minor differences of the minimum and average ISL values obtained by the three algorithms can only be observed if we check more digits after the decimal place. The major performance difference for the three algorithms lies in the consumed time and the number of conducted iterations. It can be seen from Table~\ref{Tab:ISL_T1} that the accelerated ISLNew algorithm always consumes the shortest time and needs the smallest number of iterations compared to the other two algorithms. It is capable of generating two independent waveforms of code length $ 2048 $ within $ 56 $ seconds via $ 469 $ iterations, while the CAN and accelerated ISLSong algorithms require about $ 73 $ and $ 62 $ seconds via $ 6990 $ and $ 491 $ iterations, respectively.

In addition, it can be seen from Table~\ref{Tab:ISL_T2} that the results obtained for the stopping criterion (ii) follow the same trends as in Table~\ref{Tab:ISL_T1}. It further verifies the advantages of the proposed ISLNew algorithm. All three algorithms obtain nearly the same minimum and average ISL values after convergence, which are equal to those obtained using the stopping criterion (i). Comparing Tables~\ref{Tab:ISL_T1}~and~\ref{Tab:ISL_T2}, it can also be seen that each of the three algorithms requires longer time and larger number of iterations to converge to satisfy the pre-set tolerance when the stopping criterion (ii) is used. For example, the worst consumed time and the largest number of conducted iterations have been respectively increased about $ 10 $ and $ 7 $ times for the accelerated ISLSong algorithm for code length $ P=2048 $. 

Since the three ISL minimization-based waveform design algorithms tested in this example show very minor differences in auto- and cross-correlation plots (versus time lags), i.e., all algorithms nearly achieve the lower bound of the minimum achievable ISL under both stopping criteria used, we omit to show the auto- and cross-correlation plots for the sake of brevity.

\subsection{WISL Minimization}

\begin{figure*}[!ht]
	\centering
	\subfloat[Normalized WISL values versus the number of conducted iterations. The stopping criterion (i) is used.\label{Fig:WISLVSITE_a}] { \includegraphics[width=0.464\textwidth]{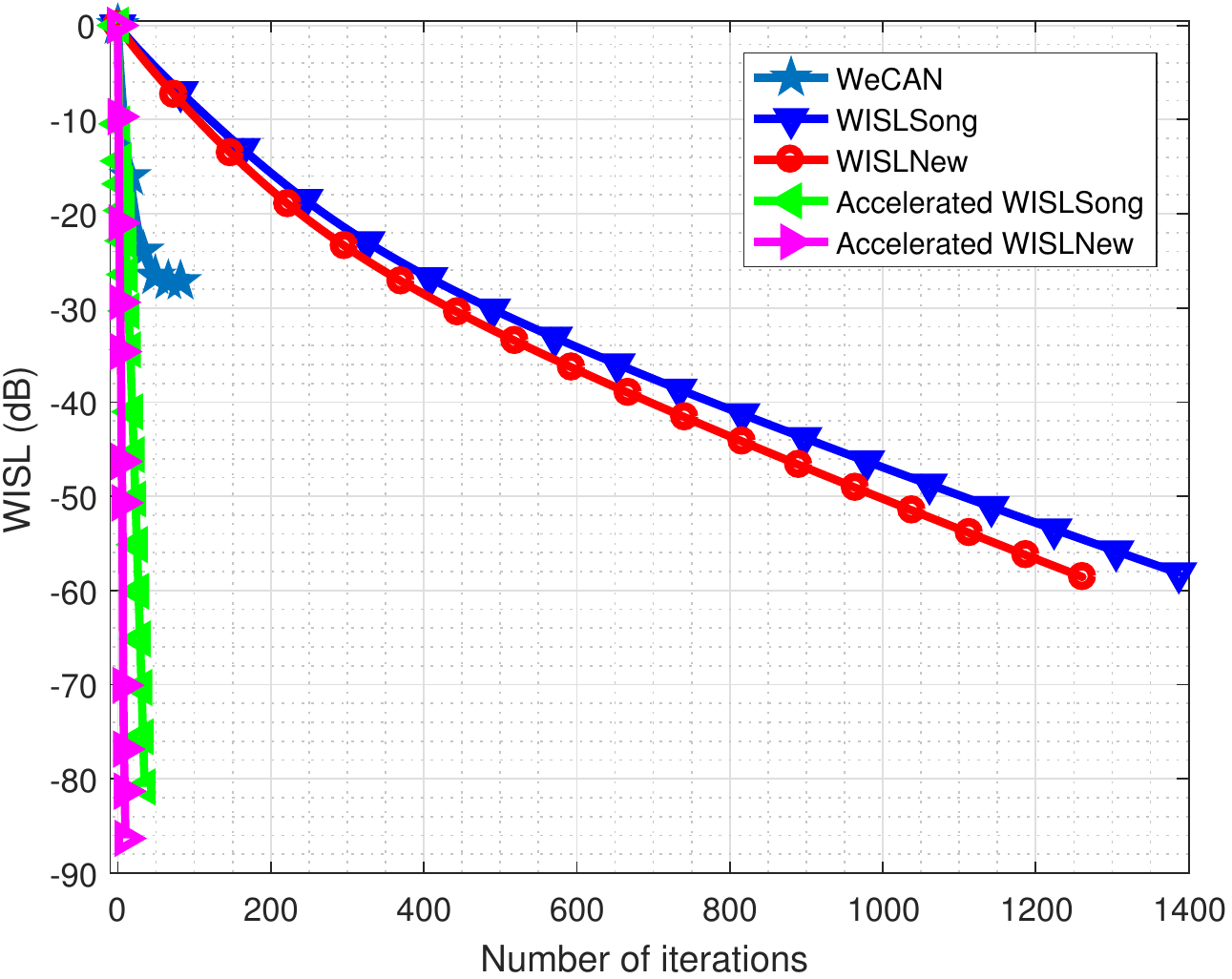} }
	\hfill
	\subfloat[Normalized WISL values versus the number of conducted iterations. The stopping criterion (ii) is used.\label{Fig:WISLVSITE_b}] { \includegraphics[width=0.45\textwidth]{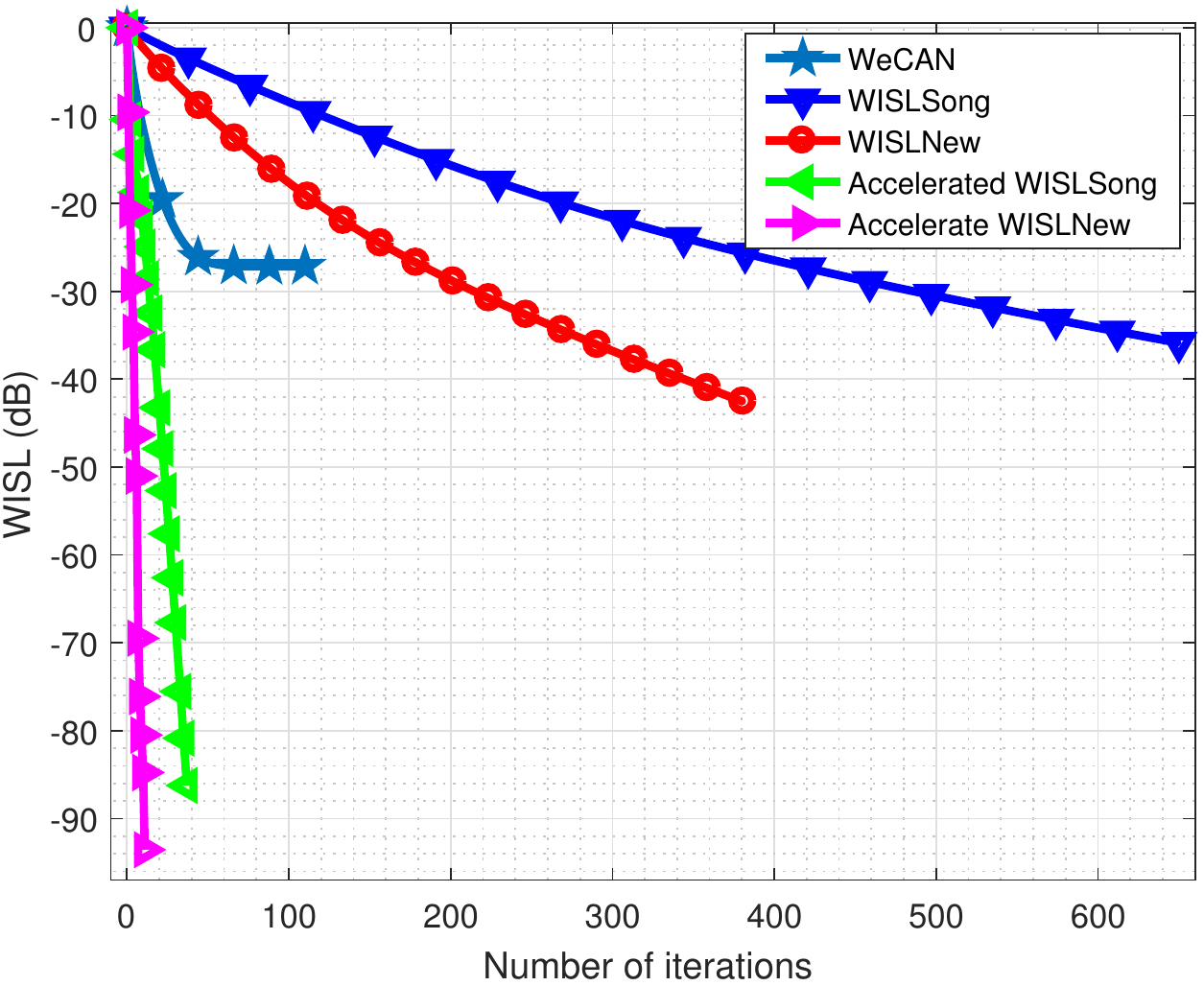} }
	\caption{Convergence of the WISL minimization-based algorithms tested. Third simulation example: $M = 2$ and $P = 256$.}
	\label{Fig:WISLVSITE}
\end{figure*}

In our third example, we study the convergence properties of the WeCAN, WISLSong, WISLNew, accelerated WISLSong, and accelerated WISLNew waveform design algorithms in terms of the number of conducted iterations for a random trial when $ M = 2 $ and $ P = 256 $. The ISL controlling weights are $  \gamma_p = 1, \, p=-19, \ldots,19 $, while the others are zeros. The WISL values obtained at each iteration of the algorithms tested are normalized by the WISL value associated with the initial set of sequences.

Similar to the first example, it can be seen from Figs.~\ref{Fig:WISLVSITE}\subref{Fig:WISLVSITE_a} and \ref{Fig:WISLVSITE}\subref{Fig:WISLVSITE_b} that the WISL values for all algorithms tested monotonically decrease as the number of iterations increases. This observation is independent of the employed stopping criteria. In this example, the accelerated WISLSong and accelerated WISLNew algorithms show much faster convergence speed and achieve significantly lower WISL values than their corresponding non-accelerated versions. The reason lies in the fact that only sidelobes associated with the time lags within $ [-19, 19] $ are controlled, which enables to achieve much lower WISL values. The WeCAN algorithm performs better than the non-accelerated WISLNew and WISLSong algorithms, however, it obtains the worst WISL after convergence. Moreover, it can be seen from both sub figures of Fig.~\ref{Fig:WISLVSITE} that the proposed WISLNew algorithm shows better convergence than the WISLSong algorithm, and its superiority is more obvious for the non-accelerated implementation. It can also be seen that the required number of iterations by the WISLSong algorithm under each of the employed stopping criteria is around two times larger than that required by the proposed WISLNew algorithm.

\begin{table*}[!th]
	\centering
	\begin{adjustbox}{max width=\textwidth}
		\begin{threeparttable}
			\caption{WISL performance comparisons (including the excess minimum and average ISL, the consumed time, and the number of conducted iterations) of the algorithms tested versus code length for $ M=2 $ waveforms for stopping criterion (i)}
			\abovecaptionskip = 7pt
			\label{Tab:WISL_T1}
			\centering
			\setlength\tabcolsep{2pt}
			\begin{tabular}{|c|c|c|c|c|c|c|c|c|c|c|c|c|c|c|c|c|c|c|c|c|c}
				\hline
				\multicolumn{1}{|c|}{\multirow{2}{*}{}} 	& \multicolumn{4}{c|} {$ P=32, M=2 $} 	& \multicolumn{4}{c|} {$ P=128, M=2 $}	& \multicolumn{4}{c|} {$ P=512, M=2 $}	& \multicolumn{4}{c|} {$ P=1024, M=2 $} 		& \multicolumn{4}{c|} {$ P=2048, M=2 $}\\ \cline{2-21}
				&  Min.\tnote{\emph{a}}\,  & Ave.\tnote{\emph{b}}\,                       &  Time                      & Iter.\tnote{\emph{c}}\, &  Min.  & Ave.                        &  Time                      & Iter. &  Min.  & Ave.                        &  Time                      & Iter. &  Min.  & Ave.                        &  Time                      & Iter. &  Min.  & Ave.                        &  Time                      & Iter. \\	\hline
				WeCAN			&	29.08		&	29.84	&	1.47		& 314	&	19.84		&	19.83	&	5.82		& 200	&	19.38		&	20.44	&	17.65		& 67 &	20.10		&	20.25	&	90.82		& 97  &	19.65		&	20.41	&	313.78		& 97 \\ \hline
				WISLSong			&	27.80		&	28.65	&	0.57		& 238	&	-34.88		&	-27.18	&	0.78		& 41 &	-47.12		&	-21.17	&	7.82		& 21	&	-40.18		&	-35.68	&	41.35		& 15		&	-48.17	&	-28.67	&	102.62		& 9	\\ \hline
				WISNew			&	27.75	&	28.44	&	1.05		& 94	&	-65.74		&	-31.76	&	0.37		& 26 &	-80.31		&	-50.37	&	2.05	& 7	&	-85.54		&	-73.83	&	6.15		& 6	&	-78.85		&	-62.46	&	20.45		& 5	\\ \hline
			\end{tabular}
			\begin{tablenotes}
				\item[\emph{a}]Min.: Obtained minimum WISL value (in dB).
				\item[\emph{b}]Ave.: Obtained average WISL value (in dB).
				\item[\emph{c}]Iter.: Number of conducted iterations.
			\end{tablenotes}
		\end{threeparttable}
	\end{adjustbox}
\end{table*}

\begin{table*}[!th]
	\centering
	\begin{adjustbox}{max width=\textwidth}
		\begin{threeparttable}
			\caption{WISL performance comparisons (including the excess minimum and average ISL, the consumed time, and the number of conducted iterations) of the algorithms tested versus code length for $ M=2 $ waveforms for stopping criterion (ii)}
			\abovecaptionskip = 7pt
			\label{Tab:WISL_T2}
			\centering
			\setlength\tabcolsep{2pt}
			\begin{tabular}{|c|c|c|c|c|c|c|c|c|c|c|c|c|c|c|c|c|c|c|c|c|c}
				\hline
				\multicolumn{1}{|c|}{\multirow{2}{*}{}} 	& \multicolumn{4}{c|} {$ P=32, M=2 $} 	& \multicolumn{4}{c|} {$ P=128, M=2 $}	& \multicolumn{4}{c|} {$ P=512, M=2 $}	& \multicolumn{4}{c|} {$ P=1024, M=2 $} 		& \multicolumn{4}{c|} {$ P=2048, M=2 $}\\ \cline{2-21}
				&  Min.\tnote{\emph{a}}\,  & Ave.\tnote{\emph{b}}\,                       &  Time                      & Iter.\tnote{\emph{c}}\, &  Min.  & Ave.                        &  Time                      & Iter. &  Min.  & Ave.                        &  Time                      & Iter. &  Min.  & Ave.                        &  Time                      & Iter. &  Min.  & Ave.                        &  Time                      & Iter. \\	\hline
				WeCAN		& 28.20	&	28.92		&	1.58	&	 374	&	18.70		&	19.61	&	7.59		& 267	&	20.72		&	20.84	&	18.47		& 84 &	20.12		&	20.17	&	56.86		& 71  &	21.24		&	21.32	&	176.46		& 66 \\ \hline
				WISLSong			&	27.48		&	28.59	&	0.84		& 298	&	-40.31		&	-39.38	&	0.80		& 51 &	-47.63		&	-46.42	&	8.55		& 14	&	-42.83		&	-39.65	&	33.24		& 20		&	-49.44	&	-48.81	&	75.34		& 16	\\ \hline
				WISLNew	& 27.67		&	28.25	&	0.30	&		149	&	-49.91		&	-37.68	&	0.20		& 24 &	-59.52		&	-55.09	&	1.22	& 7	&	-79.34		&	-74.87	&	4.15		& 6	&	-72.53		&	-61.87	&	13.79		& 5	\\ \hline
			\end{tabular}
			\begin{tablenotes}
				\item[\emph{a}]Min.: Obtained minimum WISL value (in dB).
				\item[\emph{b}]Ave.: Obtained average WISL value (in dB).
				\item[\emph{c}]Iter.: Number of conducted iterations.
			\end{tablenotes}
		\end{threeparttable}
	\end{adjustbox}
	\vspace{-10pt}
\end{table*}

In the fourth example, we compare the performance of the WISL minimization-based algorithms (WeCAN, accelerated WISLSong, and accelerated WISLNew) in terms of the following characteristics: the minimum and average WISL after convergence (in dBs), the average consumed time (in seconds), and the average number of conducted iterations. The number of waveforms and the code lengths are taken the same as in the second example, and all results are averaged over $ 50 $ independent trials. Moreover, the ISL controlling weights are the same as in the previous example. Table~\ref{Tab:WISL_T1} shows the results when the stopping criterion (i) is used, while Table~\ref{Tab:WISL_T2} shows the results for the stopping criterion (ii). 

It can be seen from Table~\ref{Tab:WISL_T1} that the accelerated WISLNew algorithm outperforms the other two algorithms tested in this example for all code lengths. The accelerated WISLSong algorithm shows the second best performance in terms of all evaluation characteristics, and the WeCAN algorithm performs the worst. Among all code lengths, the smallest average WISL values after convergence by the WeCAN, accelerated WISLSong, and accelerated WISLNew algorithms are respectively $ 19.83 $~dB, $ -35.68 $~dB, and $ -73.83 $~dB, while the largest average WISL values are $ 29.84 $~dB, $ 28.65 $~dB, and $ 28.44 $~dB (all at $ P=32 $), respectively. The WeCAN algorithm generally consumes significantly more time, requires more iterations, and achieves higher WISL values than the other two algorithms. It is manly because it does not deal with the original WISL objective function but instead a surrogate one. The advantages of the WISLNew and WISLSong algorithms over the WeCAN algorithm become a lot more obvious when the code length $ P $ is larger than $ 32 $, which verifies the fact that the WeCAN algorithm is suitable only for the WISL minimization-based waveform design with short code length. The WeCAN algorithm may converge very slowly when the set of ISL controlling weights is not sparse.

\begin{figure*}[!t]
	\centering
	\subfloat[Auto-correlation of the first designed waveform.\label{Fig:WISLCorr_a}] { \includegraphics[width=0.48\textwidth]{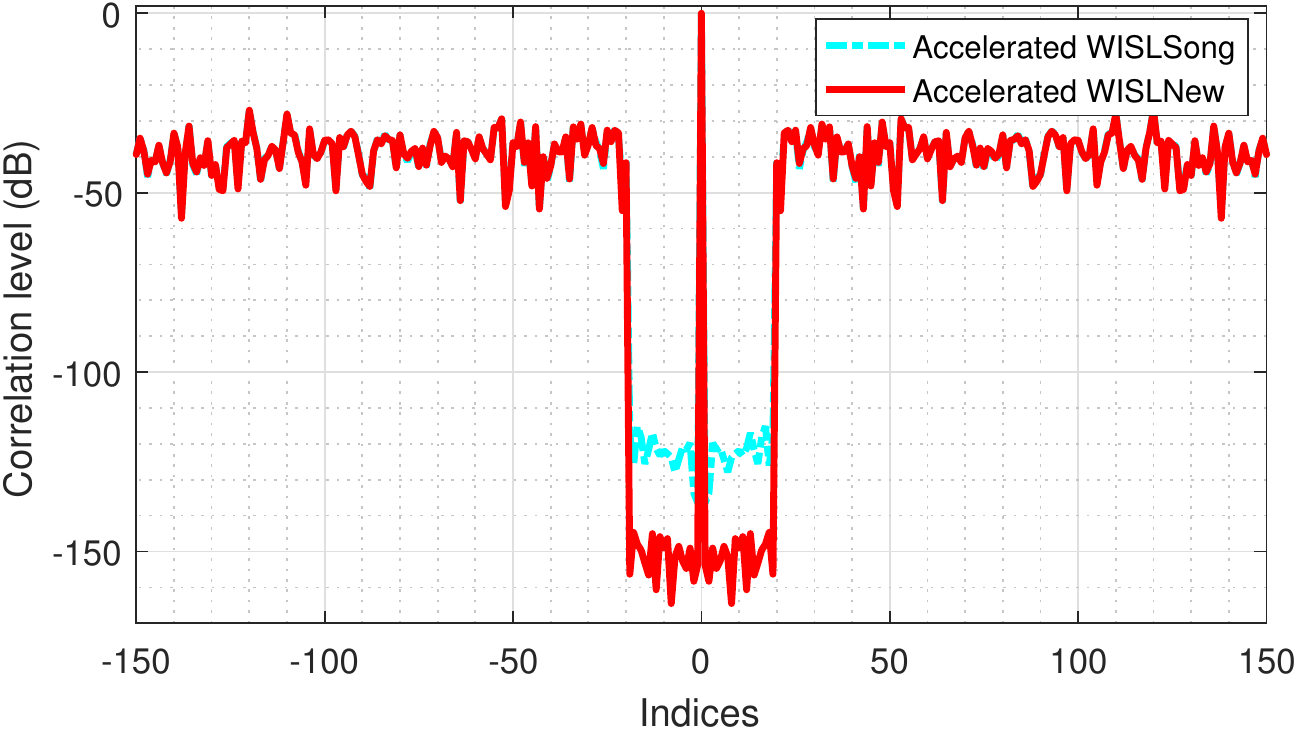} }
	\hfill
	\subfloat[Cross-correlation between the first and second designed waveforms.\label{Fig:WISLCorr_b}] { \includegraphics[width=0.48\textwidth]{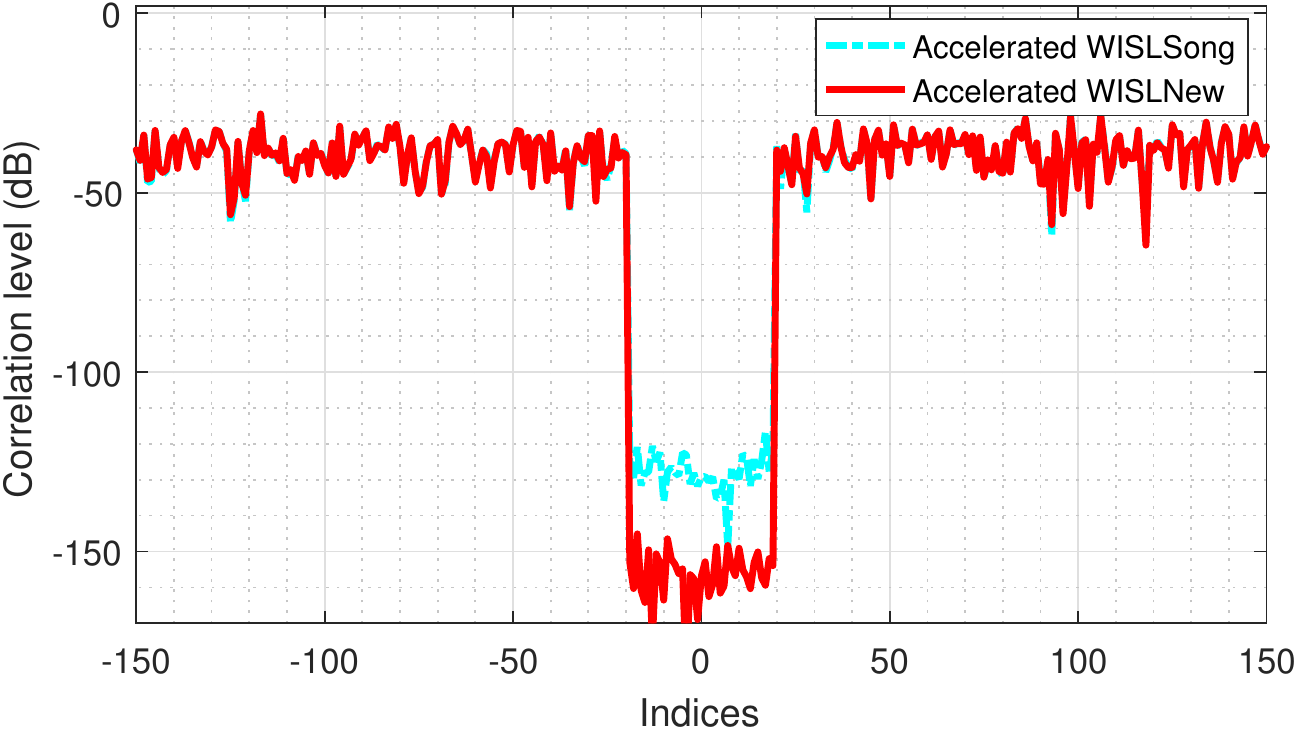} }
	\vfill
	\subfloat[Cross-correlation between the first and second designed waveforms.\label{Fig:WISLCorr_c}] { \includegraphics[width=0.48\textwidth]{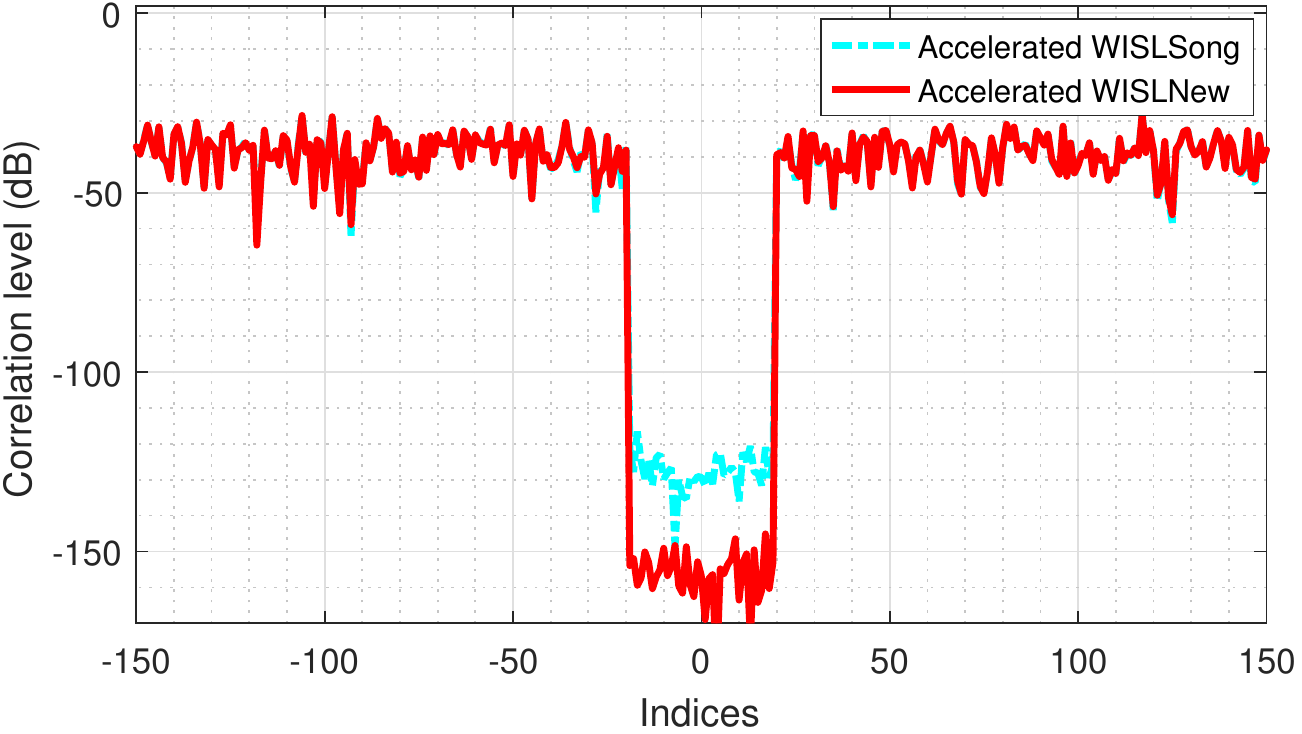} }
	\hfill
	\subfloat[Auto-correlation of the second designed waveform.\label{Fig:WISLCorr_d}] { \includegraphics[width=0.48\textwidth]{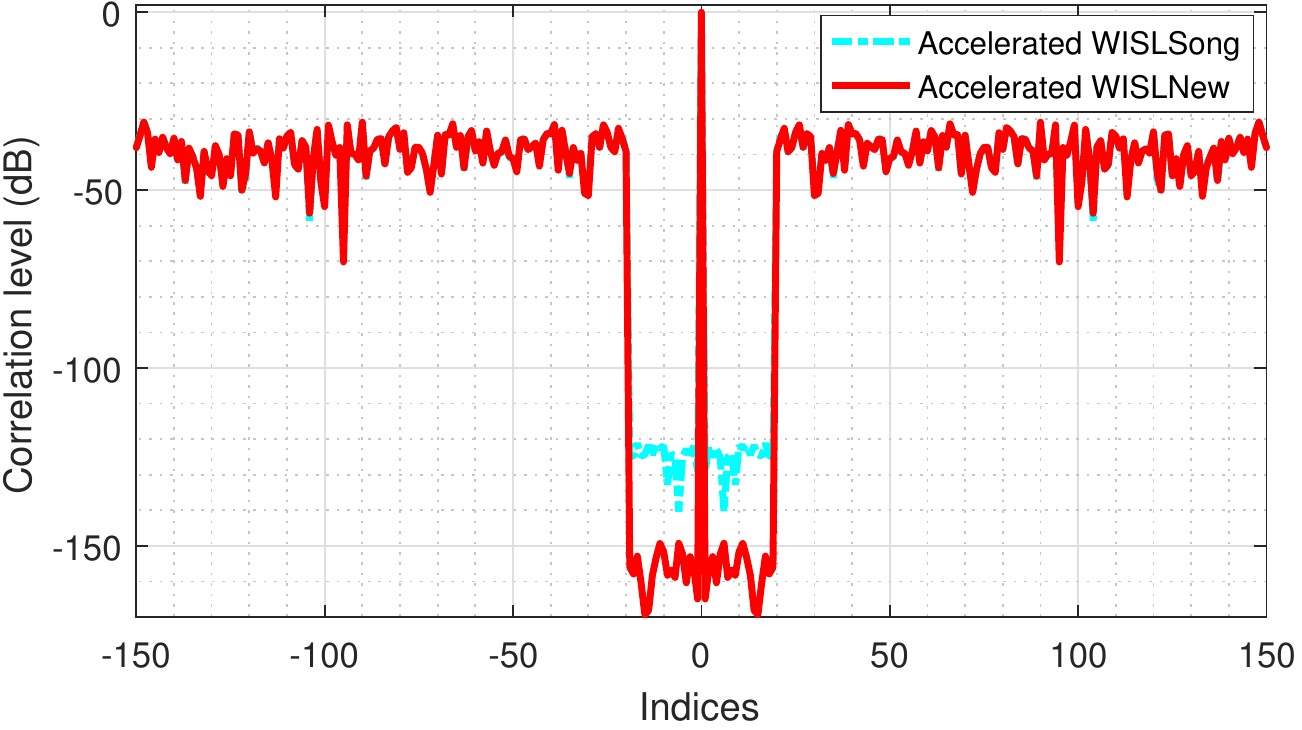} }
	\caption{Correlation property evaluation of the designed waveforms for the WISL minimization based designs. Here the number of waveforms is $ M=2 $, the large code length is $ P=4096 $, and the first stopping criterion is utilized. Correlation results with respect to time lags only within the range $ [-150, 150] $ are shown.}
	\label{Fig:CorrEvaluation}
\end{figure*}

Focusing on the comparisons between the accelerated WISLSong and accelerated WISLNew algorithms, we can see from Table~\ref{Tab:WISL_T1} that the accelerated WISLNew algorithm is superior to the accelerated WISLSong algorithm when the code length $ P $ is larger than $ 32 $, and the biggest differences (occurring at $ P = 2048 $) of the minimum and average WISL values between these two algorithms reach $ 58.24 $~dB and $ 26.53 $~dB, respectively. The accelerated WISLNew algorithm always consumes less time and requires smaller number of iterations than the accelerated WISLSong algorithm. The larger the code length $ P $ is, the more obvious the superiority of the accelerated WISLNew algorithm becomes. For example, the ratio of the consumed time between these two algorithms decreases from about $ 0.51 $ (that is, the accelerated WISLNew algorithm requires only $ 0.51 $ time required by the accelerated WISLSong algorithm) to $ 0.12 $, and the corresponding ratio of the number of conducted iterations for these two algorithms decreases from about $ 0.63 $ to $ 0.55 $ as the code length increases from $ P=128 $ to $ P = 2048 $. Thus, the proposed WISLNew algorithm is better suited for large-scale waveform design problems. In addition, for the waveform design with smaller code length and larger number of non-zero ISL controlling weights (corresponding to the $ P = 32 $ case in this example), the minimum and average WISL values after convergence for the WISLNew and WISLSong algorithms are close to each other, and are slightly better than those for the WeCAN algorithm. However, the accelerated WISLNew algorithm is still superior in terms of the other characteristics.

The above discussed advantages of the proposed WISLNew algorithm over the WeCAN and WISLSong algorithms are also verified by Table~\ref{Tab:WISL_T2} where the experiment is conducted under the stopping criterion (ii). It can be seen that the data therein follow the same trends as in Table~\ref{Tab:WISL_T1}. To be explicit, the accelerated WISLNew algorithm takes around $ 3 \thicksim 7 $ times less time and $ 2 \thicksim 3 $ times less number of iterations compared to the accelerated WISLSong algorithm, and it achieves significantly lower minimum and average WISL values after convergence, especially for large code length. On the contrary, the WeCAN algorithm always demonstrates the worst performance in terms of all evaluation characteristics. 

Finally in the fifth example, we present the auto- and cross-correlations of $  M=2 $ waveforms designed by the accelerated WISLSong and accelerated WISLNew algorithms with code length $ P = 4096 $. We set the ISL controlling weights to the same values as those in the previous two examples. The stopping criterion (i) is utilized in this example. The four sub figures in Fig.~\ref{Fig:CorrEvaluation} stand for the auto- and cross-correlations of the two sets of waveforms generated by the two aforementioned algorithms. Here the correlation levels, defined as $ 20\log_{10}(r_{mm'}(p)), \, m, m' \in \{1, 2\}; \, p \in \{-4095, 4095\}$, are shown (in dBs). To better display the results, we only show the auto- and cross-correlations for the time lags within the range $ [-150, 150] $. The WeCAN algorithm for the large code length in this example costs significantly more time and shows the worst auto- and cross-correlations, and therefore, is not shown in Fig.~\ref{Fig:CorrEvaluation}. 

It can be seen from Fig.~\ref{Fig:CorrEvaluation} that the auto-correlations associated with the time lags $ [-19, -1] \cup [1, 19]$ and cross-correlations associated with the time lags $ [-19, 19] $ for both generated sets of waveforms are well controlled, while the waveform correlations associated with other time lags are not controlled. Therefore, the latter results in much higher correlation levels. Under the condition of using the same tolerance parameter, the correlation levels corresponding to the time lags of interest obtained by the proposed WISLNew algorithm are significantly better than those obtained by the WISLSong algorithm. The largest gap between the obtained correlation levels by these two algorithms has reached about $ 30 $~dB. Moreover, the proposed WISLNew algorithm needs significantly shorter time than the WISLSong algorithm as it has been discussed above.

\section{Conclusion}\label{Sec:Conclu}
In this paper, we have developed two (one based on ISL and the other based on WISL minimization) new fast algorithms for designing single/multiple unimodular waveforms/codes with good auto- and cross-correlation or weighted correlation properties. Since the corresponding optimization problems are non-convex and may be large-scale, the proposed algorithms are based on the MaMi framework and utilize a number of newly found inherent algebraic structures in the objective functions of the corresponding optimization problems. These properties have enabled us to reduce the computational complexity of the algorithms to the level which is suitable for large-scale optimization and at least similar to or lower than that of the existing algorithms. Moreover, the proposed algorithms also show faster convergence speed to tolerance and provide waveforms of better quality than those of the existing state-of-the-art algorithms.



\end{document}